\documentclass[twocolumn, preprintnumbers, superscriptaddress,
  nofootinbib]{revtex4-1}
\pdfoutput=1
\usepackage[pdftex]{graphicx}
\usepackage{amssymb,amsmath}
\usepackage{bm}
\usepackage[colorlinks,linktocpage]{hyperref}
\hypersetup{colorlinks=true, citecolor=blue}
\usepackage[dvipsnames]{xcolor}
\begin{document}
\preprint{INR-TH-2020-045}

\title{Instability of rotating Bose stars}
\date{08/06/2021}

\author{A.S. Dmitriev}\email{dmitriev.as15@physics.msu.ru}
\affiliation{Lomonosov Moscow State University, Faculty of Physics,
  Moscow 119991, Russia}
\affiliation{Institute for Nuclear Research of the Russian Academy
  of Sciences, Moscow 117312, Russia}
\author{D.G. Levkov}\email{levkov@ms2.inr.ac.ru}
\affiliation{Institute for Nuclear Research of the Russian Academy
  of Sciences, Moscow 117312, Russia}
\affiliation{Institute for Theoretical and Mathematical Physics, MSU,
  Moscow 119991, Russia}
\author{A.G. Panin}\email{panin@ms2.inr.ac.ru}
\affiliation{Institute for Nuclear Research of the Russian Academy
  of Sciences, Moscow 117312, Russia}
\affiliation{Moscow Institute of Physics and Technology, 
  Dolgoprudny 141700, Russia}
\author{E.K. Pushnaya}\email{qutorcle@gmail.com}
\affiliation{Lomonosov Moscow State University, Faculty of Physics,
  Moscow 119991, Russia}
\affiliation{University of Cambridge, DPMMS, Cambridge CB2 1TN}
\author{I.I. Tkachev}\email{tkachev@ms2.inr.ac.ru}
\affiliation{Institute for Nuclear Research of the Russian Academy
  of Sciences, Moscow 117312, Russia}
\affiliation{Novosibirsk State University, Novosibirsk 630090,
    Russia}

\newcommand{\red}[1]{\textcolor{red}{\bf #1}}  
\newcommand{\di}[1]{\textcolor{blue}{\bf #1}}  
\newcommand{\ig}[1]{\textcolor{magenta}{\bf #1}}
\newcommand{\an}[1]{\textcolor{Green}{\bf #1}}  
\newcommand{\ek}[1]{\textcolor{Orange}{\bf #1}}

\begin{abstract}
  Light bosonic (axion--like) dark matter may form Bose stars~---
    clumps of nonrelativistic Bose--Einstein condensate supported by
    self--gravity. We study rotating Bose stars composed of
    condensed 
    particles with nonzero angular momentum $l$. We analytically
  prove that these objects are unstable at arbitrary~$l \ne 0$ if
  particle self--interactions are attractive or negligibly
  small. They decay by shedding off the particles and transporting
    the angular momentum to the periphery of the system until a
    Saturn--like configuration appears: one (or several) spin--zero
    Bose stars and clouds of diffuse particles orbit around the
    mutual center. In the case of no self--interactions we
  calculate the profiles and dominant instability modes of the
  rotating stars: numerically at $1 \leq l\leq 15$ and analytically at
  $l\gg 1$. Notably, their lifetimes are always comparable to the
  inverse binding  energies; hence, these objects cannot be considered
  long--living. Finally, we  numerically show that  in models with
  sufficiently strong repulsive self--interactions the Bose star with
  $l=1$ is stable.
\end{abstract}

\maketitle

%%%%%%%%%%%%%%%%%%%%%%%%%%%%%%%%%%%%
\section{Introduction and Main Results}
\label{sec:intro}
\begin{figure}[b]
  \centerline{\includegraphics[trim={0 0.4cm 0 0.4cm}, clip]{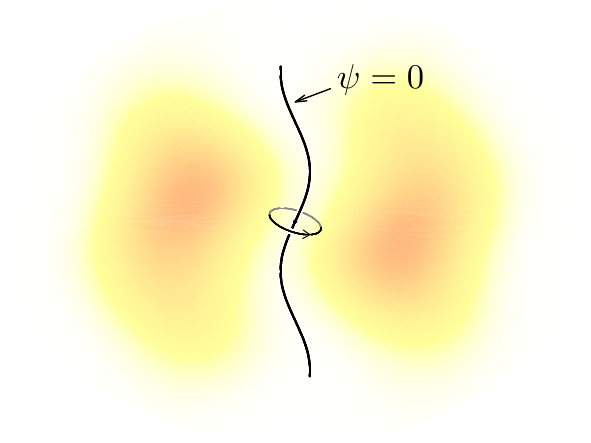}}
  \caption{(Not to scale) Bose--Einstein condensate (shaded region) rotating around
    the vortex line $\psi=0$ (solid).}
  \label{fig:vortex}
\end{figure}

Every object in the Universe can rotate around its center of mass and
carry angular momentum. There is, however, a unique substance~---
Bose--Einstein condensate of particles in a quantum state $\psi(t,\,
\bm{x})$~--- that does not rotate easily, and if does, rotates in
its own peculiar way. Indeed, the condensate velocity can be
identified~\cite{LL9} with the phase gradient divided by the
particle mass\footnote{Units with $\hbar=1$ are used in all
    dimensionful expressions.}
$$
\bm{v} = \bm{\nabla} \arg \psi(t,\, \bm{x})/m\;. 
$$
This vector is explicitly irrotational at nonzero density:
${\mathrm{rot}\, \bm{v} = 0}$ at $\psi\ne 0$. Hence, the only way to
add  rotation is to drill a hole through the condensate,
i.e.\ introduce a vortex line  $\psi=0$ in
Fig.~\ref{fig:vortex}.  And this costs energy! As a by--product, the
angular momentum of the condensate is quantized with the number
$l$ of vortex lines.
 
In the present--day Universe, the Bose--Einstein condensate of dark
matter particles may exist in the form of gravitationally self--bound
Bose stars~\cite{Kaup:1968zz, Ruffini:1969qy, Tkachev:1986tr}, 
cf.~\cite{Guth:2014hsa}. During decades, the studies of these
objects were migrating from the periphery of scientific interest
towards its focal point~\cite{Niemeyer:2019aqm}. Now, it is clear that
the Bose stars may form  abundantly by universal gravitational
mechanisms~\cite{Seidel:1993zk, Schive:2014dra, Levkov:2018kau} in the 
mainstream models with light dark matter. If the latter consists of
QCD axions, they nucleate~\cite{Levkov:2018kau,
  Eggemeier:2019jsu} inside the typical axion
miniclusters~\cite{Hogan:1988mp, Kolb:1993hw}~--- widespread
smallest--scale structures conceived at the QCD phase
transition~\cite{Kolb:1993zz, Kolb:1993hw, Vaquero:2018tib, 
  Buschmann:2019icd, Eggemeier:2019khm, Gorghetto:2020qws}. In the
case of fuzzy dark matter, gigantic Bose stars (``solitonic cores'')
appear in the centers of galaxies during structure
formation~\cite{Schive:2014dra, Schive:2014hza, Veltmaat:2018dfz}.
In both cases these objects cease growing beyond certain
mass~\cite{Schive:2014hza, Eggemeier:2019jsu, Chen:2020cef}.

\begin{figure}
  \centerline{\includegraphics{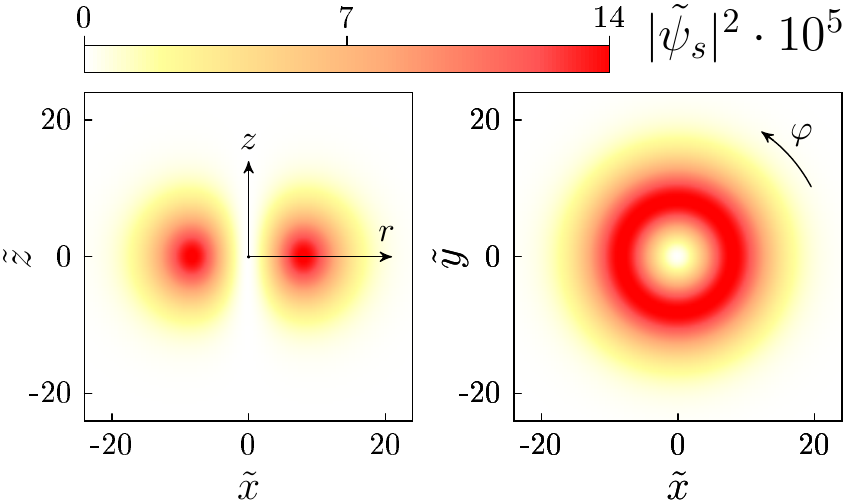}} 
  \caption{Bose star~(\ref{eq:42}) rotating with $l=1$ around
    the~$z$ axis. This configuration is computed numerically in the case of
    negligible particle self--interactions, $\lambda = 0$. Left and
    right panels show $y=0$ and $z=0$ sections of the 
    density profile $|\psi_s(\bm{x})|^2$, respectively. Tildes above
    the letters indicate dimensionless units (to be introduced in the
    main text).}
  \label{fig:bose-star}
\end{figure}

One can rotate the Bose star by drawing a vortex through its center. A
stationary and axially--symmetric Ansatz of this
type is~\cite{Silveira:1995dh, Ryan:1996nk, Schunck:1996he}  
\begin{equation}
  \label{eq:42}
  \psi_s(\bm{x}) = \psi_s(r,\, z) \, \mathrm{e}^{-i\omega_s t + il\varphi}\;,
\end{equation}
where $(r,\, \varphi,\, z)$ are the cylindrical coordinates, $\omega_s < 0$
is the binding energy of the condensed particles, and $l$ is their 
angular momentum. The total spin of the Bose star is then $J_z =
l M_s/m$, where $M_s$ is its mass. Importantly, regularity  
requires $\psi_s$ to vanish as $r^l$ at ${r\to 0}$. Thus, the symmetry axis 
of the configuration~\eqref{eq:42} is indeed a vortex line with the
winding number~$l$.

Solving numerically the coupled equations for $\psi_s$ and its
gravitational potential, we obtain the profile of the rotating Bose
star. It has  a distinctive toroidal form, see Fig.~\ref{fig:bose-star} and 
cf.~\cite{Schunck:1996he,Ryan:1996nk}.

Rotating Bose stars, if stable, would be important for astrophysics
and cosmology. Their centrifugal barriers can
resist~\cite{Davidson:2016uok, Hertzberg:2018lmt} to
bosenovas~--- collapses of overly 
massive stars due to attractive self--interactions of
bosons~\cite{Zakharov_2012, Chavanis:2011zi, Levkov:2016rkk}. This
means, in particular, that fast--rotating QCD axion stars would  reach
larger masses and densities~\cite{Hertzberg:2018zte} which may be
sufficient to ignite observable parametric
radioemission~\cite{Tkachev:1986tr, Hertzberg:2018zte, 
  Hertzberg:2020dbk, Amin:2020vja, Amin:2021tnq},
see the analytic analysis of the latter process
in~\cite{Levkov:2020txo}. Besides, the angular momenta of the Bose
stars  are detectable in principle: directly by observing gravitational
waves from their mergers~\cite{LIGOScientific:2018mvr,
  LIGOScientific:2018jsj, Abbott:2020jks} or indirectly if they
eventually collapse into spinning black holes~\cite{Kaup:1968zz,
  Ruffini:1969qy} which merge and emit gravitational waves.

Surprisingly, none of the existing simulations show nucleation of
the rotating objects~(\ref{eq:42}) from generic Cauchy
data, even if strong  spherical  asymmetry is present from the
start. We numerically observed spin--zero Bose stars
form~\cite{Seidel:1993zk, Schive:2014dra, Levkov:2018kau,
  Eggemeier:2019jsu, Chen:2020cef}, collide~\cite{Schive:2014hza},
merge~\cite{Schwabe:2016rze, Amin:2019ums, Hertzberg:2020dbk}, or
tidally disrupt~\cite{Hui:2016ltb, Du:2018qor}. In the end of the
simulations, they were strongly oscillating~\cite{Veltmaat:2018dfz,
  Marsh:2018zyw, Li:2020ryg}, random--walking~\cite{Schive:2019rrw,
  Li:2020ryg}, partially or completely destroyed~\cite{Schive:2014hza, 
  Du:2018qor}, but never acquired a nonzero angular  momentum. In
addition, relativistic cousins of rotating Bose stars~--- the
lumps of complex scalar field with $U(1)$ conserved charge bounded by
Einstein gravity~--- were numerically shown to develop remarkable
axially asymmetric instabilities~\cite{Sanchis-Gual:2019ljs,
  DiGiovanni:2020ror, Siemonsen:2020hcg}.

All of this strongly suggests that rotating nonrelativistic Bose stars
are unstable, although skeptics still may argue that some of the
observed numerical instabilities could be artificially inflicted by
the Cartesian lattices breaking axial symmetry. Besides,
relativistic Bose stars~--- even at zero spin~--- have
essentially different stability properties~\cite{Tkachev:1986tr,
  Visinelli:2017ooc} in phenomenologically interesting cases of QCD
axions and axion--like particles as compared to the models with global
$U(1)$ symmetry~\cite{Sanchis-Gual:2019ljs, DiGiovanni:2020ror, 
  Siemonsen:2020hcg}.

%%%%%%%%%%%%%%%%%%%%%%%%%%%%%%%%%%%%%%%
\subsection*{Executive summary}
Let us summarize the main results and approaches leaving their derivation
and technical details to the main text.

In this paper we analytically prove a no--go theorem: nonrelativistic
gravitationally bound Bose stars~(\ref{eq:42}) with arbitrary nonzero
angular momentum are unstable in models with negligibly small
($\lambda=0$) or attractive ($\lambda < 0$) particle
self--interactions. This result is applicable in the popular cases 
of fuzzy and QCD axion dark matter.  On the other hand, in models with
repulsive self--interactions ($\lambda > 0$) a stability region for
  the $l=1$ Bose star exists.

\begin{figure}
\centerline{\includegraphics{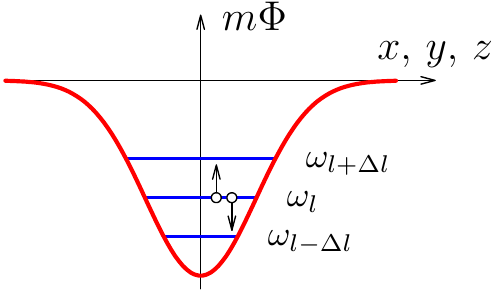}}
\caption{(Not to scale) Instability of the rotating Bose star.}
\label{fig:transition}
\end{figure}

Our approach reveals the mechanism for the instability: it is caused
by the pairwise transitions of the condensed particles from the
original state with the angular momentum $l$ to the $l+\Delta l$  and
$l - \Delta l$ states, see Fig.~\ref{fig:transition}. This process
conserves the total spin and decreases the potential energy
of the Bose star. Piling up due to Bose factors, the particle
  transitions lead to exponential growth of the axially asymmetric
perturbations:
\begin{equation}
  \label{eq:17}
  \delta \psi \propto \mathrm{e}^{\mu t}\;,
\end{equation}
where $\mu$ is the complex exponent and $(\mathrm{Re}\, \mu)^{-1}$ is
the lifetime of the rotating configuration~(\ref{eq:42}). 

\begin{figure}
  \includegraphics{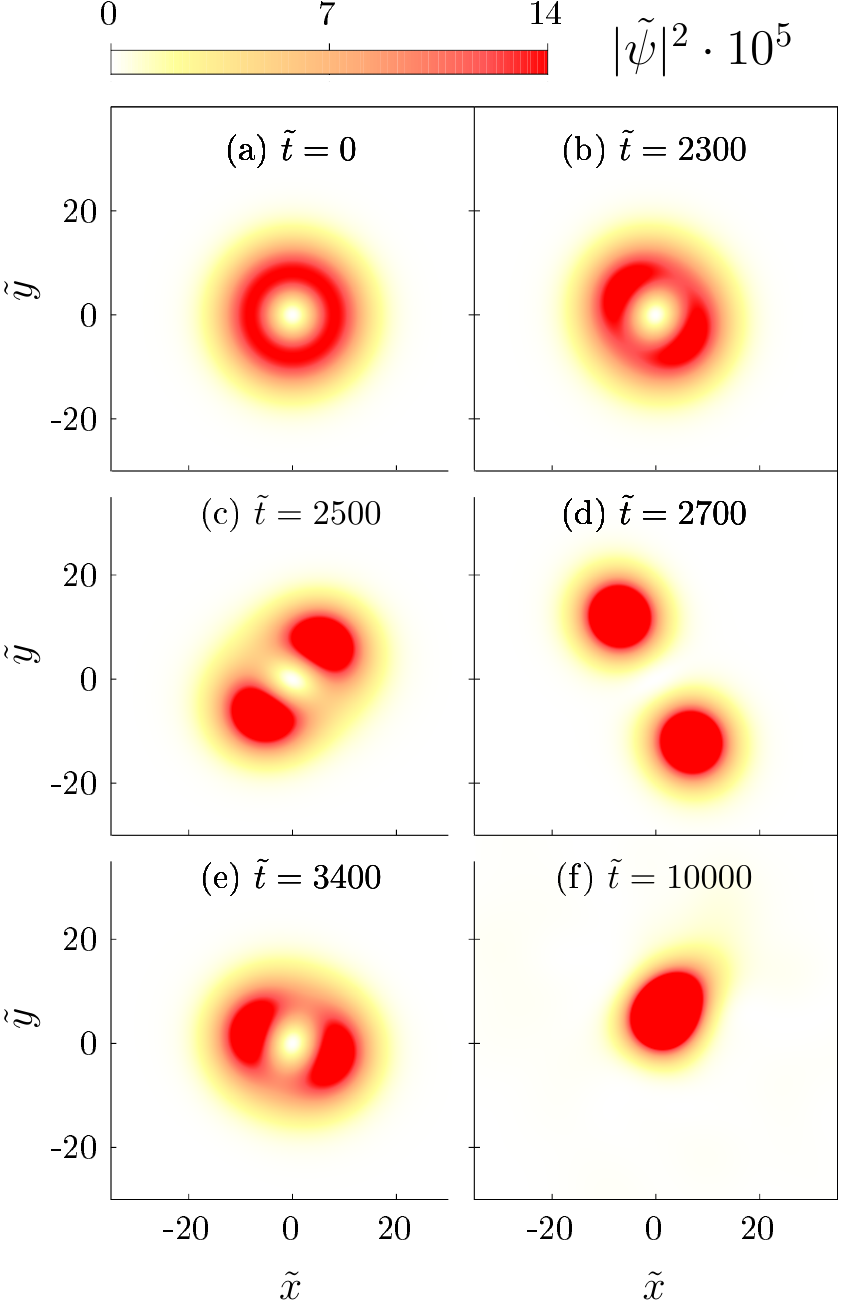}
  \caption{Numerical evolution of the $l=1$ Bose star in the case of
    purely gravitational interactions $\lambda=0$, see~[\onlinecite{movie}a]
    for the related movie. Frames (a)---(f) display horizontal (${z=0}$)
    sections of the density $|\psi|^2$ at different times. Units
    are specified in the main text.}
  \label{fig:inst-comic}
\end{figure}

In Fig.~\ref{fig:inst-comic} we visualize three--dimensional
 numerical evolution of the perturbed $l=1$ Bose star, see
also the movie~[\onlinecite{movie}a] and cf.~\cite{Sanchis-Gual:2019ljs,
  DiGiovanni:2020ror, Siemonsen:2020hcg}. The panels (a)---(f)
  display horizontal sections of the solution at fixed time moments.
The simulation starts in  Fig.~\ref{fig:inst-comic}a with the star
profile distorted by an invisibly small asymmetric
perturbation $\delta   \psi \sim 10^{-6}\psi_s$. The  latter
grows exponentially with time, becomes discernible at the moment
of  Fig.~\ref{fig:inst-comic}b and reaches a fully nonlinear regime
$\delta \psi \sim \psi_s$ in Fig.~\ref{fig:inst-comic}c. At this
point, a bound system of two spherical Bose stars appears. They
oscillate and rotate around the mutual center of mass in
Figs.~\ref{fig:inst-comic}c---e. Finally, one of the stars gets
tidally disrupted, whereas the other survives. The evolution ends in
Fig.~\ref{fig:inst-comic}f  with nonspinning Bose star surrounded by a
cloud of diffuse axions. They rotate around the mutual center of mass.

\begin{figure}
\centerline{\includegraphics{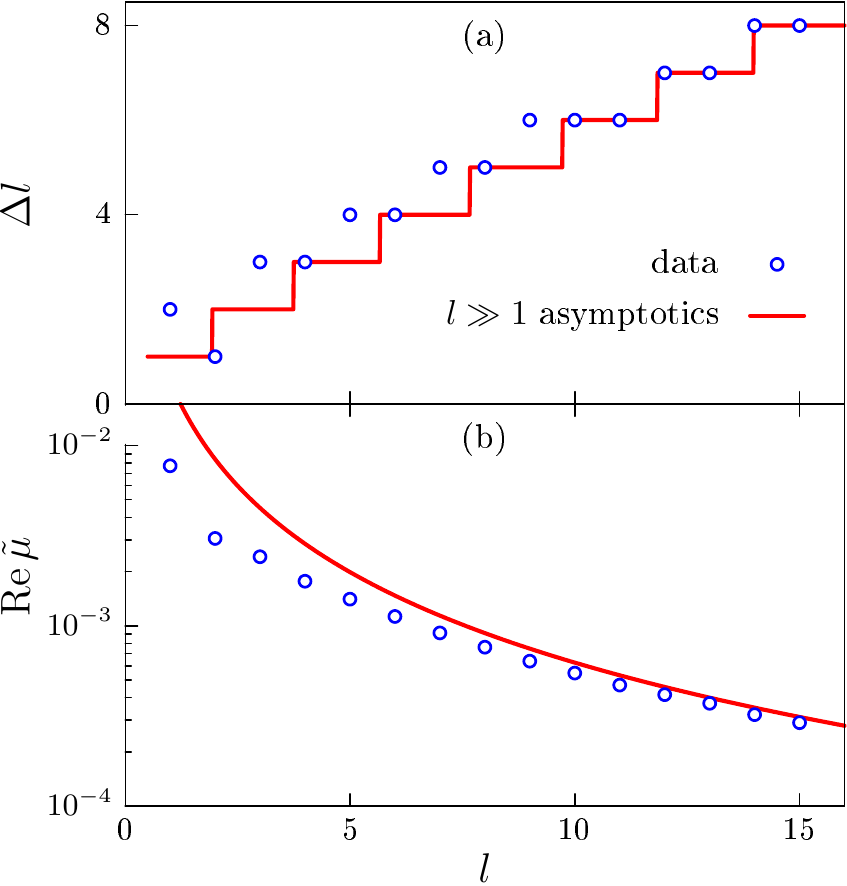}}
\caption{Parameters of the dominant instability modes in the
  backgrounds of Bose stars with different $l$: (a) angular momentum
  transfers $\Delta l$ and (b) growth exponents $\mathrm{Re} \, 
  \mu$ in units of Eq.~(\ref{eq:9}). We consider negligible
  self--interaction of particles, $\lambda=0$. Numerical data  
  (circles) rapidly approach large $l$ asymptotics~(\ref{eq:5}) (lines).}
\label{fig:instab-exponent}
\end{figure}

We explicitly compute the dominant instability modes of the rotating Bose
stars in the case of purely gravitational interactions ($\lambda=0$):
numerically at moderately small $l$  and analytically at $l \gg 1$. We also
identify the angular momentum transfers $\Delta l$ in the respective
particle transitions. In physical units, the complex growth exponents
of the instability modes have the form,  
\begin{equation} 
  \label{eq:9}
  \mu = \tilde{\mu}\,m^3G^2M_s^2\;,
\end{equation}
where the dimensionless parameter $\tilde{\mu}$ and integer
$\Delta l$ depend only on~$l$. Their numerical values are listed in
Table~\ref{tab:parameters_instability} and displayed in
Fig.~\ref{fig:instab-exponent} (circles). 

\begin{table}
  \begin{tabular}{cccc|cccc}
\hline
$l$ & $\Delta l$ & ~$\mathrm{Re}\, \tilde{\mu} \cdot 10^3$ &
~$\mathrm{Im}\, \tilde{\mu} \cdot 10^3$ & 
$l$ & $\Delta l$ & ~$\mathrm{Re}\, \tilde{\mu}\cdot 10^3$ &
~$\mathrm{Im}\, \tilde{\mu} \cdot 10^3$ \\
\hline
1 & 2 & $7.73$  & $-16.2$~ &
2 & 1 & $3.05$  & $-9.64$\\
3 & 3 & $2.42$  & $-6.82$~ &
%4 & 3 & $1.77$  & $-4.13$ 
5 & 4  & $1.41$  & $-3.45$\\
7 & 5  & $0.91$  & $-2.12$~&
10 & 6 & $0.55$ & $-1.13$\\
15 & 8 & $0.29$ & $-0.58$~&
$\gg 1$ & \multicolumn{2}{c}{Eq.~\eqref{eq:5}}\\
\hline
\end{tabular}
  \caption{Parameters of the dominant instabilities in
    rotating Bose stars with different $l$: angular momentum transfers
    $\Delta l$ and complex growth exponents $\mu$, see
    Eqs.~(\ref{eq:17}), (\ref{eq:9}). The  case of purely
    gravitational interactions ($\lambda=0$) is considered.}
\label{tab:parameters_instability}
\end{table}

At $l\gg 1$ and $\lambda=0$ the parameters $\Delta l$ and
$\tilde{\mu}$ can be obtained analytically:
\begin{equation}
  \label{eq:5}
  \begin{array}{l}
    \Delta l \approx  \left[0.944  \cdot l / \sqrt{\alpha_l}\right]\;,\\[.5em]
    \tilde{\mu} \approx (2.22 - 2.39 \, i\, \sqrt{\alpha_l} ) \cdot
    10^{-2} \, \alpha_l / l^2\;,
    \end{array} \qquad (l\gg 1)
\end{equation}
with corrections suppressed by $l^{-1}$. Here $[\cdot]$ denotes the 
closest integer and order--one parameters $\alpha_l$ satisfy
nonlinear equation\footnote{With numerical solution $\alpha_l \approx
  \{1.02,\, 1.51,\, 1.82,$ $2.05,\, 2.23,\, 2.38,\, $ $2.51,\, 2.62,\,
  2.72,\, 2.81\}$  at $l = \{1,\, \dots,\, 10\}$ and large $l$
  asymptotics ${\alpha_l  = \ln l} + O(\ln\ln l)$.}
\begin{equation}
  \label{eq:71}
  2 \alpha_l + 3/2 + \ln (\beta \alpha_l/l^2) = 0 \;\;\;\; \mbox{with}
  \;\;\;\; \beta \approx 2.86\cdot 10^{-2}\,. 
\end{equation}
Figure~\ref{fig:instab-exponent} shows that Eqs.~(\ref{eq:5}) (lines)
approach the numerical data at high angular momenta, though
at crude level they are already valid at $l\sim 1$.

The analytic method giving Eqs.~(\ref{eq:5}) is based on a simple
observation that the $l\gg1$ Bose stars have forms of parametrically
thin rings with cross--section profiles satisfying a set of ordinary
differential equations. We believe that this technique can be generalized to highly
nontrivial situations, in particular, to nonzero self--coupling and to
the relativistic model of~\cite{Sanchis-Gual:2019ljs,
  DiGiovanni:2020ror, Siemonsen:2020hcg}.   

Notably, in model with ultralight (fuzzy) dark matter the lifetimes
of unstable rotating Bose stars can formally exceed the age of the
Universe. Indeed, in this case Eqs.~(\ref{eq:9}), (\ref{eq:5}) give, 
\begin{equation}
  \notag
  (\mathrm{Re}\, \mu)^{-1} \simeq 10^{10}\, \mbox{yr} \cdot l^2 \,
  \left(\frac{m}{10^{-22} \, \mathrm{eV}}\right)^{-3}
  \left(\frac{M_s}{4\cdot 10^{7} \, M_{\odot}}\right)^{-2},
\end{equation}
where we used $\alpha_l\sim 1$. One may hastily conclude that these
configurations are stable on the cosmological timescales if
$m\sim 10^{-22}\, \mathrm{eV}$, $l\gtrsim 1 $, 
and $M_{s}\lesssim 4\cdot10^7\, M_\odot$. But in fact, their lifetimes
are always comparable to the oscillation periods $2\pi / \omega_s$ in
Eq.~(\ref{eq:42}): 
\begin{equation}
  \label{eq:15}
  |\omega_s|/\mathrm{Re}\, \mu \approx 1.7 \, (\alpha_l+1) \sim O(1)\;,
\end{equation}
where the large $l$ analytics was used, again. Thus, these Bose stars are
not the long--living composite objects, as their particles cannot be
assigned to the fixed--energy levels ${\omega = \omega_s}$ of the
  Bose--Einstein condensates at timescales of order or smaller
  than $|\omega_s|^{-1}$. In addition, the lifetimes of these rotating
objects are shorter than the free--fall times $t_{\mathrm{free}}
\sim l/|\omega_s|$ in their gravitational fields and hence much
smaller than their nucleation times in reasonable formation
mechanisms~\cite{Seidel:1993zk, Schive:2014dra, 
  Levkov:2018kau, Eggemeier:2019jsu, Eggemeier:2019jsu,
  Chen:2020cef}. All of this leaves only one way to observe the
rotating Bose stars in simulations: tune the initial data to their
profiles with exponential precision, like we did in Fig.~\ref{fig:inst-comic}.

\begin{figure}
  \centerline{\includegraphics{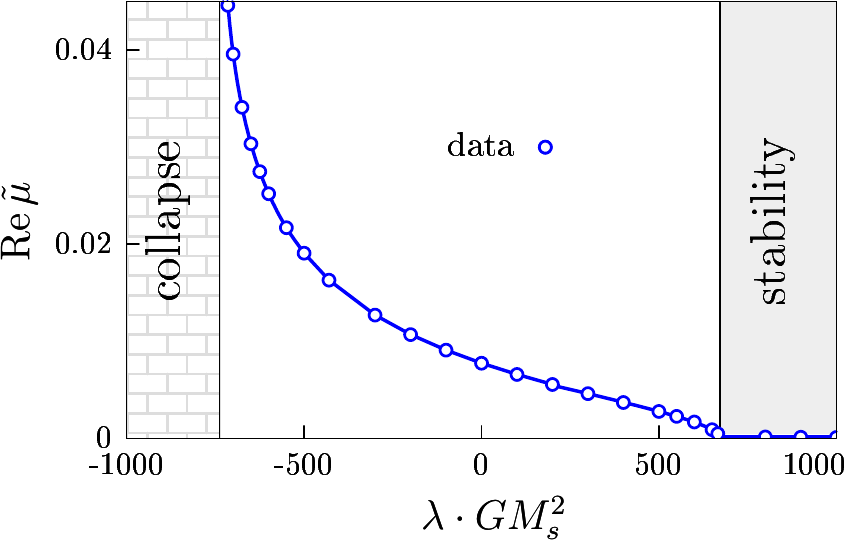}}
  \caption{The fastest instability exponent $\mathrm{Re}\, \mu$ of the
    $l=1$ Bose star as a function of the boson self--coupling
    $\lambda$ (line--points) at $M_s = \mbox{const}$. Units of
    $\mu$ are introduced in Eq.~\eqref{eq:9}. Shaded regions
    correspond to collapsing stars at large negative $\lambda$
    and absolutely stable stars at $\lambda > \lambda_0$.}
  \label{fig:mu_lambda} 
\end{figure}

Finally, we investigate numerically (in)stability of the $l=1$
Bose star in the model with nonzero particle self--interactions.
Figure~\ref{fig:mu_lambda} shows the result of this study:
dependence of the dominant decay exponent  $\mathrm{Re}\, \mu$ on the
self--coupling~$\lambda$ at a given star mass $M_s$ (points and an
interpolating line). Notably, the stationary fixed--mass Bose
stars with $l=1$ exist only at ${\lambda > \lambda_{cr}}$,
where\footnote{Errorbars in Eqs.~(\ref{eq:25}), (\ref{eq:20}) are
    related to numerical errors. They estimate the sensitivities of fits
    for $\lambda_{cr}$ and $\lambda_0$ to lattice parameters, see  
    Sec.~\ref{sec:self-inter-cond} for details.} 
    \begin{equation}
    \label{eq:25}
    \lambda_{cr} =(-738 \pm 4) / (GM_s^2)\;.
  \end{equation}
In the case of
stronger attraction\footnote{Or, conversely, at masses above critical
  $M_{cr}^{(l=1)} \approx 27.2/(-\lambda G)^{1/2}$ in models with
  $\lambda<0$.  As anticipated in~\cite{Hertzberg:2018zte, Hertzberg:2018lmt},
  this critical mass exceeds the respective value at $l=0$,
    cf.~\cite{Chavanis:2011zi,  Levkov:2016rkk}.}  (smaller $\lambda$) these objects
collapse~\cite{Chavanis:2011zi}, i.e.\ 
squeeze towards the higher density regions in a
self--similar fashion ~\cite{Zakharov_2012, Levkov:2016rkk}.
This process is shown in the movie~[\onlinecite{movie}b].

Figure~\ref{fig:mu_lambda} confirms that the $l=1$ Bose stars are
indeed unstable in the cases of negligibly weak or attractive
self--interactions:  $\mathrm{Re}\, \mu > 0$  at $\lambda \leq 
0$. This result is in agreement with our no--go theorem. In addition,
we see that the instability region extends towards moderately small 
positive couplings $\lambda < \lambda_0$, where 
  \begin{equation}
    \label{eq:20}
    \lambda_0 \approx (672 \pm 2) / (GM_s^2)\;
  \end{equation}
marks the beginning of the gray region in Fig.~\ref{fig:mu_lambda}.
  
  At $\lambda > \lambda_0$, however, our numerical data  are
    consistent\footnote{More precisely, the numerical procedure
      formally gives ${\mathrm{Re}\, \tilde{\mu} < 2\cdot 10^{-4}}$
      which is below the accuracy of calculations.} with
    $\mathrm{Re}\, \mu  \approx 0$. This suggests that the $l=1$ star is
    absolutely stable\footnote{When our work was 
  already completed, a numerical investigation of rotating
  Bose--Einstein condensed galaxy halos with repulsive particle
  self--interactions has appeared~\cite{Nikolaieva:2021owc}. The parameters of that
    study correspond to $\lambda \approx 6\,\lambda_0$~--- hence, a
  stable vortex with $l=1$ was observed.} at large $\lambda$, see the
    movie~[\onlinecite{movie}c]. Conversely, this   happens at large
    mass, ${M_s \gtrsim 
  25.9/(G\lambda)^{1/2}}$. Thus, the  
Bose--Einstein condensate behaves more like a solid body if
self--repulsion dominates in the equations. The same
stabilization effect was previously observed in the relativistic case 
in~\cite{Siemonsen:2020hcg}.

This paper is organized as follows. We introduce rotating Bose stars
in Sec.~\ref{sec:rotating-bose-stars}, prove no--go theorem
on their instability in Sec.~\ref{sec:no-go-theorem}, and numerically 
illustrate decay of  the $l=1$ star in
Sec.~\ref{sec:numer-illustr}. Then we compute the instability modes of
all rotating objects: numerically at moderately small $l$ in  
Sec.~\ref{sec:instability-modes} and analytically at $l\gg 1$
in Sec.~\ref{sec:analytic-solution-at}. In Sec.~\ref{sec:discussion}
we discuss generalizations of our results.

%%%%%%%%%%%%%%%%%%%%%%%%%%%%%%%%%%%%%%%
\section{Rotating Bose stars}
\label{sec:rotating-bose-stars}
A system of self--gravitating nonrelativistic bosons is described at
large occupation numbers by collective wave function $\psi(t,\,
\bm{x})$ and gravitational potential $\Phi(t,\,\bm{x})$. The latter
quantities satisfy Gross--Pitaevskii and Poisson equations,
\begin{align}
  \label{eq:12}
  & i\partial_t \psi = - \frac{\Delta \psi}{2m} + \left( m\Phi +
  \frac{\lambda |\psi|^2}{8m^2}\right)\psi\;,\\
   \label{eq:13}
    &\Delta \Phi = 4 \pi m G |\psi|^2\;,
\end{align}
where the extra potential $\lambda |\psi|^2/8m^2$ represents
contact self--interaction of bosons with
self--coupling\footnote{Entering as $\lambda a^4/4!$ into the scalar
  potential of a relativistic field~$a(t,\,\bm{x})$.}~$\lambda$.
Below we consider all three cases of repulsive, attractive, and
  negligible self--interactions: $ \lambda>0 $, $ \lambda<0 $, and
$\lambda =0$. The last two are especially
interesting,  as they are relevant for the popular dark matter models with
QCD axions~\cite{diCortona:2015ldu} and ultralight (fuzzy)
axion--like particles~\cite{Schive:2014dra, Niemeyer:2019aqm}.

One can define the rotating  Bose star as a stationary and
axially--symmetric solution~(\ref{eq:42}) to the system
(\ref{eq:12}), (\ref{eq:13}). This means that its wave function
$\psi_s(r,\, z)$ obeys a stationary Scr\"odinger equation
\begin{equation}  
  \label{eq:45}
  \omega_s\, \psi_s = - \frac{\Delta \psi_s}{2m} + \left( m\Phi_s +
  \frac{\lambda |\psi_s|^2}{8m^2}\right)\psi_s\;,
\end{equation}
whereas $\Phi = \Phi_s(r,\, z)$   satisfies
  Eq.~(\ref{eq:13}). Hereafter we mark all Bose star 
quantities with the subindex~$s$ and  keep in mind that 
axially--symmetric Laplacian ${\Delta \psi_s \equiv \partial_z^2\psi_s
  + r^{-1} \partial_r
  (r\partial_r \psi_s) - l^2\psi_s/r^2}$ includes a centrifugal barrier
in the last term. Apparently, Eq.~(\ref{eq:45}) describes
nonrelativistic particles occupying a single 
level of their self--made potential well ${ {m} \Phi_s +
  \lambda |\psi_s|^2/8m^2}$. All of them have energy $\omega_s$ and
angular momentum $l$.

Note that $m$ and $G$ disappear from all equations after coordinate
and field rescalings with arbitrary parameter $v_0$: ${\bm{x} =
  \tilde{\bm{x}}/mv_0}$, $t  = \tilde{t}/mv_0^2$ or 
${\omega_s = mv_0^2\, \tilde{\omega}_s}$, ${\psi = v_0^2(m/G)^{1/2}}
\tilde{\psi}$, and ${\Phi = v_0^2 \tilde{\Phi}}$. We perform numerical
calculations in these dimensionless units selecting $v_0 = mGM$ to
make the total rescaled mass equal to one: $\tilde{M} = 1$. This
leaves only one constant in the equations: the rescaled self--coupling  
$\tilde{\lambda} = \lambda GM^2$ replacing~$\lambda$.

Solving Eqs.~(\ref{eq:13}), (\ref{eq:45}), one obtains Bose stars
at different $l$ and $\tilde{\lambda}$; we will gradually
introduce relevant numerical and analytic techniques in
Secs.~\ref{sec:numer-illustr}---\ref{sec:analytic-solution-at}. All
rotating (${l\geq 1}$) objects have distinctive toroidal forms, like 
the one with $l=1$ and $\lambda=0$ in Fig.~\ref{fig:bose-star}. 

\begin{table}
  \begin{tabular}{c c|cc|cc}
    $l$ & $\tilde{E}_s \cdot 10^3$ & $l$ & $\tilde{E}_s\cdot 10^3$ &
    $l$ & $\tilde{E}_s \cdot 10^3$ \\
    \hline
    $0$~ & $-54.2$~ &
    $1$~ & $-19.0$~ &
    $2$~ & $-10.3$ \\
    $3$~ & $-6.57$~ &
    $4$~ & $-4.64$~ &
    $5$~ & $-3.49$ \\
    $6$~ & $-2.74$~ &
    $7$~ & $-2.24$~ &
    $10$~ & $-1.34$\\\cline{5-5}\rule{0pt}{4ex}
    $15$~ & $-0.736$~ &
    ~$\gg 1$~ & \multicolumn{3}{l}{$\tilde{E_s} \approx
      -\alpha_l(\alpha_l+1)/(8\pi^2 l^2)$} \\[2ex] 
    \hline
\end{tabular}
  \label{tab:bs_parameters}
  \caption{Energies of the rotating Bose stars at different $l$ in the
      case $\lambda = 0$; physical units can be restored using
      Eq.~(\ref{eq:21}). The data with $l\leq 15$ are obtained
      numerically, while the last item lists large $l$
      asymptotics of Sec.~\ref{sec:instability-modes}. Parameters
      $\alpha_l$ obey Eq.~(\ref{eq:71}).}
\end{table}

To investigate (in)stability of these stars, we need another kind of
analysis. It is important that the nonrelativistic
evolution~(\ref{eq:12}), (\ref{eq:13}) conserves a number of
quantities: the total mass $M$ and multiplicity $N$ of bosons,
\begin{equation}
  \label{eq:46}
  M \equiv mN =  m\int d^3 \bm{x} \,  |\psi|^2\;,
\end{equation}
their energy
\begin{equation}
  \label{eq:56}
  E = \int d^3 \bm{x} \left[ \frac{|\bm{\nabla}\psi|^2}{2m} +
    \frac{m}2\, \Phi |\psi|^2 + \frac{\lambda |\psi|^4}{16 m^2} \right]\;,
\end{equation}
and the components of the net angular momentum, e.g.\
\begin{equation}
  \label{eq:57}
  J_z = -i \int d^3 \bm{x} \, \psi^*\partial_\varphi \psi\;,
\end{equation}
where $\varphi = \mathrm{arctan} (y/x)$ is the angular cylindrical coordinate.
The values of these integrals characterize the Bose stars. Say,
the $l=1$ star in Fig.~\ref{fig:bose-star} has ${\tilde{E}_s \approx
  -0.019}$ or, in physical units, 
\begin{equation}
  \label{eq:21}
  E_s = \tilde{E}_s\, m^2 G^2 M_s^3\;, 
\end{equation}
where we performed rescaling in Eq.~(\ref{eq:56}). The energies 
of some  Bose stars at $\lambda=0$ are listed in
Table~\ref{tab:bs_parameters}. Their total spins are proportional
to the masses: $J_{z,\, s} = l M_s/m$, cf.\ Eqs.~(\ref{eq:42}) and~(\ref{eq:57}). 

Now, observe that the Bose stars, rotating or not, extremize the total
energy $E$ at a given mass $M = M_s$. In other words, they are the extrema
of the functional ${F \equiv E - \omega_s N}$ where the Lagrange
  multiplier $\omega_s$ fixes ${N = M/m}$. To show this
  explicitly, we rewrite the energy~(\ref{eq:56}) in the form
\begin{multline}
  \label{eq:43}
  E = \int d^3 \bm{x} \Bigg[ \frac{|\bm{\nabla}\psi|^2}{2m} +
    \left(m\Phi + \lambda u \right) |\psi|^2 \\
    + \frac{(\bm{\nabla} \Phi)^2}{8\pi G} - 4\lambda m^2u^2
    \Bigg]
\end{multline}
including the gravitational and self--interaction potentials
$\Phi(\bm{x})$ and $u(\bm{x})$. Once this is done,
the functional $F$ reaches extremum at $\Phi = \Phi_s$ and
$u=u_s$ satisfying
\begin{equation}
  \label{eq:47}
  \Delta \Phi = 4 \pi G m |\psi|^2  \;\;  \mbox{and}\;\; u = 
   |\psi|^2 / 8m^2\;.
\end{equation}
One can substitute this solution back into Eq.~(\ref{eq:43}) and
recover the old energy expression (\ref{eq:56}). Further variation of $F$
with  respect to $\psi^*(\bm{x})$ gives 
the Gross--Pitaevskii equation~(\ref{eq:45}) which together with
Eqs.~(\ref{eq:47}) forms the same stationary system for the Bose  star
profile as before. Thus, rotating Bose  stars are indeed the
extrema of $F$ labeled with $l$ and~$M_s$.

The question is whether these objects are the  local minima
  of energy at a fixed mass $M_s$ and total spin $J_{z,\, s}$. In the
next Section we will show that at $l\geq 1$ and $\lambda\leq 0$ they
are not. Rather, they are the energy saddle points which can be
destroyed by an arbitrarily small perturbation growing
exponentially with time.

It is worth noting that the above argument identifies $\omega_s$
with the binding energy of particles inside the Bose star,
cf.\ Eq.~(\ref{eq:42}). Indeed,  
infinitesimally small number of extra particles changes the energy
$E_s$ and number $N_s$ of bosons, but not the value of $F$ which is
extremal. Thus,
\begin{equation}
  \label{eq:53}
  dE_s =\omega_s dN_s\;,
\end{equation}
i.e.\ every new particle comes in with energy $\omega_s$. At
${\lambda = 0}$ this last relation can be combined with
Eq.~(\ref{eq:21}) to give
\begin{equation}
  \label{eq:2}
  \omega_s = 3\tilde{E}_s\, m^3 G^2 M_s^2 = 3mE_s/M_s\;,
\end{equation}
which is useful for numerical tests. 

%%%%%%%%%%%%%%%%%%%%%%%%%%%%%%%%%%%%%%%%%%%%%
\section{No--go theorem at $\lambda \leq 0$}
\label{sec:no-go-theorem}
Let us prove that rotating Bose stars~(\ref{eq:42})
are unstable at $l \ne 0$ if the self--coupling of their particles is
negligible or attractive, ${\lambda \leq 0}$. These cases are
  special because at $\lambda \leq 0$ the new energy functional
  (\ref{eq:43}) reaches {\it minimum} with respect to $\Phi$ and $u$ at
  their physical values~(\ref{eq:47}). Thus, we can consider generic
  independent variations of $\psi(\bm{x})$, $\Phi(\bm{x})$, and
  $u(\bm{x})$. The Bose star will be proved unstable if one of
  such variations decreases the energy~(\ref{eq:43}), since physical
  variation with $\delta\Phi$ and $\delta u$
  provided by Eqs.~(\ref{eq:47}) decreases the energy even
  further.

We introduce an auxiliary tool: a set of 
wave functions $\Psi_{l'}(\bm{x}) \propto \mathrm{e}^{il' \varphi}$ with
angular momenta $l'$ satisfying the Schr\"odinger equation in 
the  Bose star potential~(\ref{eq:47}),
\begin{equation}
  \label{eq:48}
  \omega_{l'} \Psi_{l'} = -\frac{\Delta \Psi_{l'}}{2m} + (m\Phi_s +
  \lambda u_s) \, \Psi_{l'}\;.
\end{equation}
For every $l'$ we select the eigenfunction with the minimal $\omega_{l'}$ and
normalize it to unity: $\int d^3 \bm{x} \, |\Psi_{l'}|^2 = 1$.

Notably, $\Psi_{l'}$  are not the vibrational modes of the Bose star: the
latter include related perturbations of $\psi$, $\Phi$, and $u$.
But Eq.~(\ref{eq:48}) brings in simple quantum mechanical logic 
which  will be useful in what follows. First, at ${l'}=l$ this equation coincides
with Eq.~(\ref{eq:45}) for the condensate  profile. Thus,
$\omega_l \leq \omega_s$, where strict inequality corresponds to
the case of radially excited condensate. Second, and as
a consequence of the  first, the eigenvalues  $\omega_{l'}$ with $l' < l$
are lower than $\omega_s$ by a margin because they have weaker
centrifugal barriers ${l'}^2/2mr^2$. In particular,
Eqs.~(\ref{eq:48}) with ${l'}=l$ and $l'=0$ give,
\begin{equation}
  \label{eq:49}
  \omega_s - \omega_0 \geq \int d^3 \bm{x} \; \frac{ l^2 \,
    |\Psi_l|^2}{2mr^2} > 0\;,
\end{equation}
where we observed that $\omega_0$ is the minimal eigenvalue of the
radial Hamiltonian, i.e. the operator in Eq.~(\ref{eq:48})
without the $\varphi$ derivatives. Third, in the limit ${l'} \to \infty$
the eigenfunctions $\Psi_{l'}$ become 
large in size and therefore interact only with the $\mbox{large--}\bm{x}$
asymptotics of the potential $\Phi_s \to -GM/|\bm{x}|$, but not with
its short--range part $u_s$. The respective
eigenvalues resemble the ones of the Hydrogen atom: ${\omega_{l'}
  \approx -m^3 G^2  M^2({l'}+1)^{-2}/2 \sim O(l'^{-2})}$ at~${l' \gg 1}$.  

Now, let us explicitly construct an infinitesimally small
deformation that decreases the energy (\ref{eq:43}) of the
original Bose star configuration $\{\psi_s',\, \Phi_s',\, u_s'\}$ 
  with multiplicity $N_s'$ and winding number $l\geq 1$. First, we
  extract $dN_s$ particles from the condensate thus obtaining
the star $\{\psi_s,\, \Phi_s,\, u_s\}$ with $N_s = N_s'  - dN_s$ particles 
and the same winding number. Second, we add back $dN_0$ particles in
the non--rotating state $\Psi_0$ and  $dN_{l'}$ particles in the state
$\Psi_{l'}$ with $l'\gg 1$. This process does not modify the total mass
and spin if
\begin{equation}
  \label{eq:50}
  dN_s = dN_0 + dN_{l'} \qquad \mbox{and} \qquad
  l \, dN_s  = {l'}\,  dN_{l'}\;.
\end{equation}
Physically, such deformation corresponds to a simultaneous transition of $dN_s$
condensate particles  from the state with angular momentum  $l$ to 
$l'=0$ and $l'\gg 1$ states.

At the level of configurations, we infinitesimally deform $\Phi_s'$,
$u_s'$ to the potentials $\Phi_s$, $u_s$ of the smaller--mass Bose
star and change
\begin{equation}
  \label{eq:51}
  \psi_s' \to \psi = \psi_s(\bm{x}) + dN_0^{1/2} \, \Psi_0(\bm{x}) +
  dN_{l'}^{1/2} \, \Psi_{l'} (\bm{x})\,.
\end{equation}
Substituting Eq.~(\ref{eq:51}) into the expression~(\ref{eq:43}), we
obtain the potential energy $E_f \equiv E[\psi,\,
  \Phi_s,\, u_s]$ of this final state:
\begin{equation}
  \label{eq:52}
  E_f = E_s + \omega_0 dN_0 + \omega_{l'} dN_{l'} \;,
\end{equation}
 where $E_s$ is the energy of the Bose star with $N_s$ particles and we
used Eq.~(\ref{eq:48}) for $\Psi_0$ and $\Psi_{l'}$. The cross--terms
between $\psi_s$, $\Psi_0$, and $\Psi_{l'}$ vanish due to different
dependences on $\varphi$: recall that $\psi_l \propto
\mathrm{e}^{il\varphi}$ and ${\Psi_{l'}\propto \mathrm{e}^{i{l'}\varphi}}$,
whereas $\Psi_0$, $\Phi_s$, and $u_s$ are $\varphi$--independent. 

On the other hand, we started from the Bose star with $N_s+dN_s$
particles and energy
\begin{equation}
  \label{eq:54}
  E_s' = E_s + \omega_s dN_s + O(dN_s^2)\;,
\end{equation}
see Eq.~(\ref{eq:53}). Thus, change of the potential
energy in the above transition equals
\begin{equation}
  \label{eq:55}
  E_f - E_s' = (\omega_0 - \omega_s) dN_s + O({l'}^{-1})\, dN_s < 0\;,
\end{equation}
where $dN_0$ and $dN_{l'}$ were expressed from Eqs.~(\ref{eq:50}) and we
recalled that $\omega_{l'} = O({l'}^{-2})$. The last inequality follows from 
Eq.~(\ref{eq:49}). We conclude that the deformation (\ref{eq:51})
indeed decreases the potential energy of the rotating Bose star.  

The above argument proves that all rotating Bose stars are unstable at
$\lambda \leq 0$ and arbitrary $l\geq 1$. It also qualitatively identifies
the instability mechanism. Namely, the potential energy of the 
rotating Bose star decreases if some particles perform transitions to
nonrotating states and give their angular momenta to other
particles going to the periphery of the system. A presumable
end--state of this process is a Saturn--like configuration: one or
several spin--zero Bose stars surrounded by a rotating cloud of
diffuse particles.

A warning is in order. So far we considered an explicit but very
non--optimal way of decreasing the Bose star energy. In  particular,
we voluntarily deformed the potentials $\Phi_s$, $u_s$ and fixed
the angular momenta of the particle end--states. We will see below  
that the fastest--growing modes represent pairwise  transitions of
the condensed bosons to the states with angular momenta $l\pm \Delta l$,
see Fig.~\ref{fig:transition}.

%%%%%%%%%%%%%%%%%%%%%%%%%%%%%%%%%%%%%%%%%%%%%
\section{Decay of  the $l=1$ Bose star}
\label{sec:numer-illustr}
Now, we explicitly visualize the instability of the $l=1$ Bose star
in the model with $\lambda=0$. We introduce periodic 
  spatial lattice $\{x_n,\, y_m,\, z_k\}  = \{ n \delta,\, m \delta
,\, k\delta \}$ with uniform spacing $\delta$ and fields $\psi_{n,\, m,\,   k}
\equiv \psi(x_n,\, y_{m},\, z_{k})$, $\Phi_{n,\, m,\, k}$ sitting on
its sites. Since the lattice breaks the rotational symmetry, we will
be extra cautious in separating discretization effects from 
the physical rotational instabilities.

We observe that our cubic lattice is invariant with respect to
the $\pi/2$
rotations $\hat{R}_{\pi/2}$ around $z$ axis which map
lattice points $(x_n,\, y_m)$ to $(-y_m, \, x_n)$ leaving $z_k$  
unchanged. From the technical viewpoint, this means that the time
  evolution in Eqs.~\eqref{eq:12}, \eqref{eq:13} commutes with
$\hat{R}_{\pi/2}$ even after discretization.  On the other hand, the  
fixed--$l$ configurations~(\ref{eq:42}) are the eigenvalues of these
rotations,
\begin{equation}
  \label{eq:59}
  \hat{R}_{\pi/2}\,  \psi_{n,\, m,\, k} \equiv \psi_{-m,\, n,\, k} =
  \mathrm{e}^{i \pi l/2}\, \psi_{n,\, m,\, k}\;.
\end{equation}
Thus, we define the lattice version of the $l=1$ Bose star as a
minimum--energy configuration satisfying Eq.~(\ref{eq:59}) with
the eigenvalue $\mathrm{e}^{i\pi/2}$. Such configuration is
  readily produced by the Euclidean relaxation procedure
summarized in Appendix~\ref{sec:numerical-methods}. At the end of
the  relaxation, the discretized equations~(\ref{eq:13}),
  (\ref{eq:45}) are solved almost exactly, up to negligible round--off 
errors. The solution is shown in Fig.~\ref{fig:bose-star}. 

It is worth noting that the Bose stars with ${l = 1+4k}$ have the same
eigenvalue in Eq.~(\ref{eq:59}) and cannot be separated from the $l=1$
star on this basis. But their  centrifugal barriers are essentially
stronger; hence, energy minimization still selects the configuration
with $l=1$. To the contrary, the Bose stars with $l= 0$  
and $l=-1$ have the same or 
smaller energies, but they are excluded by Eq.~(\ref{eq:59}).

Once the Bose star $\psi = \psi_s(\bm{x})$ with $l=1$ is
found, we perturb it by adding an asymmetric perturbation,
\begin{equation}
  \label{eq:58}
  \psi = \psi_s(\bm{x}) +  A\,  \mathrm{e}^{-(r/r_1)^2}
  \left[\frac{r}{r_1}\, \cos \varphi + \frac{r^2}{r_1^2}\, \cos 
    2\varphi\right]
\end{equation}
where $\tilde{r}_1 = 10$ and $\tilde{A}  = 10^{-8}$  is tiny. Then,
evolving Eqs.~(\ref{eq:12}) and (\ref{eq:13}), we watch the
star fall apart.  A numerical method for that is described in
Appendix~\ref{sec:numerical-methods}. 

The result is shown Fig.~\ref{fig:inst-comic}, see also the
  movie~[\onlinecite{movie}a]. The Bose star   remains stationary and
toroidal at first. But then it splits into two spherical
objects rotating around the mutual center of
mass. With time, one of the objects persists and the other 
gets tidally disrupted. The final configuration includes a
non--rotating Bose star surrounded by a cloud of diffuse
particles.

We stress that the $l=1$ Bose star is destroyed by the
perturbation~(\ref{eq:58}) growing exponentially in its  
background, not by something else. Without this kick and
the round--off errors, it would remain stationary, as its energy
  is minimal in the sector with fixed $\hat{R}_{\pi/2}$  and the
latter operator commutes with the time evolution. We checked that at
$A=0$ the rotating star falls apart at much larger time scales, since
initial perturbations in this case are provided by 
the round--off errors.

To quantify the growing instabilities, we subtract the original
Bose star from the numerical solution $\psi(t,\, \bm{x})$ and then
split the residual into four parts belonging to the sectors with
different eigenvalues of the $\pi/2$ rotations: 
\begin{equation}
  \label{eq:60}
  \psi(t,\, \bm{x}) = \psi_s(\bm{x})\,  \mathrm{e}^{-i\omega_s t}  +
  \psi_{0} + \psi_{1} + \psi_{2} + \psi_{3}\;.
\end{equation}
Here $\psi_{l'}(t,\, \bm{x})$ satisfy $\hat{R}_{\pi/2}\psi_{l'} =
\mathrm{e}^{i\pi l'/2} \psi_{l'}$ at every time~$t$. Roughly
speaking, they have angular momenta ${l' = 0\div 3}$, although
higher $l'$ contributions are also possible. In
Appendix~\ref{sec:numerical-methods} we construct an explicit 
projector for the decomposition~(\ref{eq:60}). 

\begin{figure}
  \centerline{\includegraphics{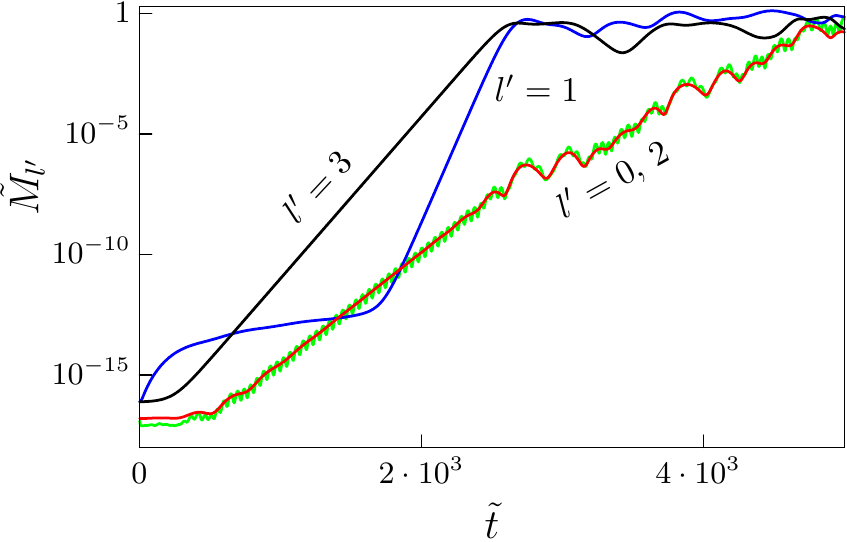}}
  \caption{Norms $M_{l'}(t)$ of the instability modes growing
    exponentially in the background of the $l=1$ Bose  star. Tildes
    indicate dimensionless  units; $\lambda=0$.}
  \label{fig:delta-psi-evolution}
\end{figure}
The norms $M_{l'}(t) \equiv m\int d^3 \bm{x} \, |\psi_{l'}|^2$ of
  the perturbations $\psi_{l'}$ are plotted in
  Fig.~\ref{fig:delta-psi-evolution}. The graphs with ${l' = 0,\, 2,\,  3}$
  grow exponentially indicating that their modes are proportional
  to ${\psi_{l'} \propto  \mathrm{e}^{\mu_{l'} t}}$ at the linear stage
  $\tilde{t}\lesssim 2500$. The perturbation with $l'=3$ is dominant:  
  it grows faster and becomes nonlinear first. Fitting its graph with
  the exponent, we obtain ${\mathrm{Re}\, \tilde{\mu}_{3} \approx 7.74
    \cdot 10^{-3}}$ in agreement with the value
    in Table~\ref{tab:parameters_instability} which is provided by the
    axially symmetric method of the next Section. The respective 
  angular momentum transfer equals ${\Delta l  \equiv |l'-l| = 2}$ modulo~4.

The subdominant graphs in Fig.~\ref{fig:delta-psi-evolution}
  deserve two remarks. First, the norms of  $\psi_{0}$ and
  $\psi_{2}$ are comparable at all times and $\mu_0  = \mu_2$. We will
  see below that these perturbations satisfy coupled linear equations
  and therefore describe the same instability with $\Delta l = |l'-l|
  = 1$. Second, the graph with $l'=1$ remains bounded at first and
  then starts growing with the exponent $\mathrm{Re}\, \mu_{1} \approx
  2 \mathrm{Re}\, \mu_{3}$. This is because the respective
    perturbation belongs to the same $l=1$ sector as the Bose star itself
    and cannot grow at the linear level. At later times, however,
    the dominant mode $\psi_{3}$ becomes large and starts sourcing
    $\psi_1\propto |\psi_3|^2$ via nonlinear terms in the equations.  

Since the dominant instability of the $l=1$ star develops with
$\Delta l=|l'-l| = 2$, the respective density perturbation  has
maxima at two angles: $|\psi_s\,\mathrm{e}^{i\varphi} + \psi_{3}|^2 -
|\psi_s|^2 \propto \cos(2\varphi + \mathrm{const})$. That is why the
background star splits into two pieces in
Fig.~\ref{fig:inst-comic}c. The subsequent nonlinear evolution in
Figs.~\ref{fig:delta-psi-evolution}d-f is intricate, however,
because the original $l=1$ star has lower energy than the two
  isolated nonrotating objects with masses $M_s/2$, see
  Eq.~(\ref{eq:21}) and Table~\ref{tab:bs_parameters}. As a consequence,
the two half--mass stars remain bound to each other until the
perturbations $\psi_{0}$ and $\psi_{2}$ grow to nonlinearity,
too. Once this happens, the bound state gets broken and the
final nonrotating object forms, see Fig.~\ref{fig:inst-comic}f. 

%%%%%%%%%%%%%%%%%%%%%%%%%%%%%%%%%%%%%%%%%%%%%
\section{Linear instabilities at arbitrary $l$}
\label{sec:instability-modes}

%%%%%%%%%%%%%%%%%%%%%%%%%%%%%%%%%%%%%%%%%%%%%
\subsection{No self--interaction}
\label{sec:no-self-interaction}

The numerical method of Sec.~\ref{sec:numer-illustr} is ideal for
visualizing the instability of the $l=1$ Bose star but it is also
not applicable at higher spins. Indeed, fast--rotating
objects cannot be separated from their lower~$l$ brothers by $\pi/2$
rotations and energy minimization. But nevertheless, we want to
compute their instability modes and complex exponents
$\mu$. We start with the case of zero
self--coupling,~$\lambda=0$.

\begin{figure}
  \centerline{\includegraphics{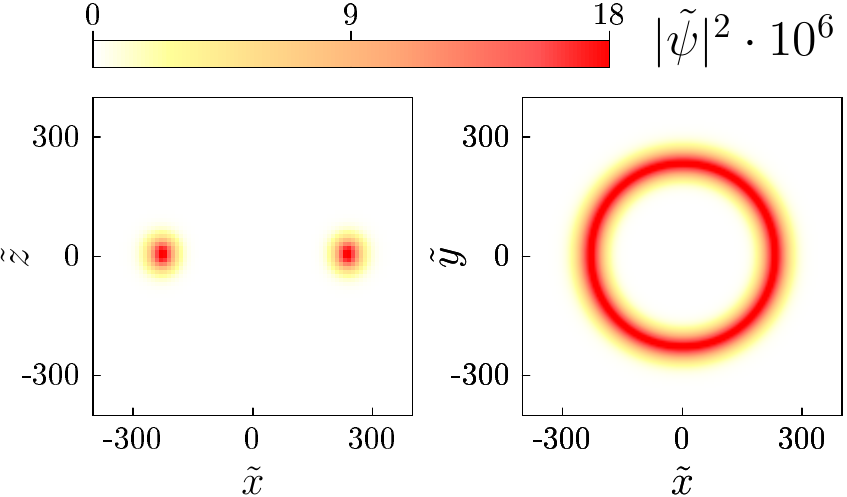}}
  \caption{Rotating Bose star with $l=10$
    and~$\lambda=0$. Dimensionless units of
    Sec.~\ref{sec:rotating-bose-stars} are used.}  
  \label{fig:large-l-star}
\end{figure}

We compute the stationary profiles $\psi_s(r,\, z)$, $\Phi_s(r,\,
z)$ of the higher~$l$ stars by solving the axially symmetric
system~(\ref{eq:12}), (\ref{eq:13}). To this end we alternate
Euclidean time steps for the field $\psi$ with successive
overrelaxation (SOR) sweeps for the gravitational potential
$\Phi$, see Appendix~\ref{sec:axially-symm-simul} for details. The
numerical procedure converges to minimal energy configurations
with given~$l$~-- rotating Bose stars. In practice, we use it
at moderately large $l = 1\div 15$. All obtained
  solutions\footnote{Of course, the profile and parameters of the
  $l=1$ Bose star coincide with  the ones computed in
  Sec.~\ref{sec:numer-illustr}.} have distinctive toroidal forms, see 
Fig.~\ref{fig:large-l-star}.

\begin{figure}
  \centerline{\includegraphics{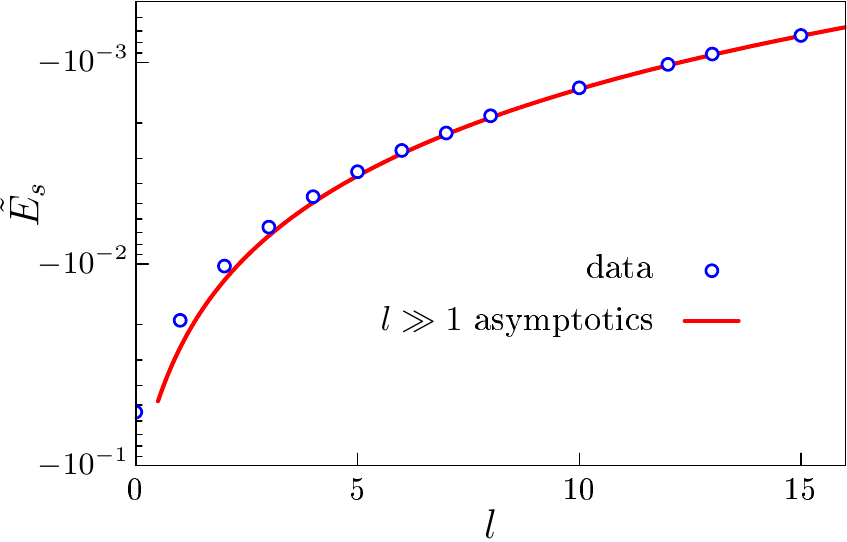}}
  \caption{Energies $\tilde{E}_s$ of the rotating Bose stars at zero 
    boson self--coupling $\lambda=0$; dimensionless units are
    introduced in Eq.~(\ref{eq:21}). Points represent numerical
    data of Sec.~\ref{sec:instability-modes}, and the solid line
    is the analytic large $l$ result of
    Sec.~\ref{sec:analytic-solution-at}.}
  \label{fig:energy_l}
\end{figure}

The energies of the Bose stars with different $l$ are listed
  in Table~\ref{tab:bs_parameters} and visualized in
  Fig.~\ref{fig:energy_l} (points). At large $l$ they approach the
  analytic expression (line) which will be derived in
  the next Section.

Next, we perturb the Bose stars to question their stability. A
  generic perturbation of $\psi$ has arbitrary dependence on
    $\varphi$ and therefore includes modes with arbitrary angular 
  momenta~$l'$. One can see, however, that at the linear level the
  modes with $l' =l +  \Delta l$ and $l' = l - \Delta l$  couple
  to each other but not to other modes. Thus, every such pair can
  be considered independently, and we write: 
\begin{align}
  \notag
  & \psi = \left[ \psi_s(r,\, z) + \delta \psi\, \mathrm{e}^{i\Delta l \varphi}
  + \delta \bar{\psi}^* \,\mathrm{e}^{-i\Delta l \varphi}\right]
  \mathrm{e}^{-i\omega_s t + il \varphi}\;,\\
  \label{eq:3}
  & \Phi = \Phi_s(r,\, z) + \delta \Phi\, \mathrm{e}^{i\Delta l \varphi} +
  \delta \Phi^* \mathrm{e}^{-i\Delta l \varphi}\;,
\end{align}
where $\delta \psi$, $\delta \bar{\psi}$, and $\delta \Phi$ depend
only on $r$, $z$, and $t$. 

Substituting Eq.~(\ref{eq:3}) into the Schr\"odinger--Poisson system
(\ref{eq:12}), (\ref{eq:13}), we arrive to equations, 
\begin{align}
& (\omega_s + i\partial_t) \delta \psi =  - \Delta_{r,z}
  \delta \psi/2m + m\psi_s \delta \Phi  \notag \\
  \notag
& \qquad \qquad\;\;
  + \left[(l+\Delta
    l)^2/(2mr^2) + m\Phi_s \right]  \delta \psi\;, \\
& (\omega_s - i\partial_t) \delta \bar{\psi} =  - \Delta_{r,z}
  \delta \bar{\psi}/2m + m\psi_s^* \delta \Phi \notag \\
  \label{eq:64}
& \qquad \qquad\;\;
    + \left[(l-\Delta l)^2/(2mr^2) + m\Phi_s \right]  \delta
    \bar{\psi}\;, \\
    \notag
& \Delta_{r,z} \delta \Phi - \Delta l^2 \delta \Phi / r^2 = 4\pi Gm
    \left(\psi_s^* \delta \psi + \psi_s \delta \bar{\psi} \right)\;,
\end{align}
where nonlinear terms in $\delta \psi$, $\delta \bar{\psi}$, 
$\delta \Phi$ are omitted, and $\Delta_{r,\, z} \equiv 
\partial_r^2  + r^{-1}\partial_r + \partial_z^2$ is the
  radial part of the Laplacian.  The last line in 
Eqs.~(\ref{eq:64}) includes both $\delta \psi$ and $\delta
\bar{\psi}$, so they are not independent, indeed. This feature
explains, in particular, why the modes with $l' = 0$ and 2 grow with 
  the same exponent in Fig.~\ref{fig:delta-psi-evolution}.

\begin{figure}
  \centerline{\includegraphics{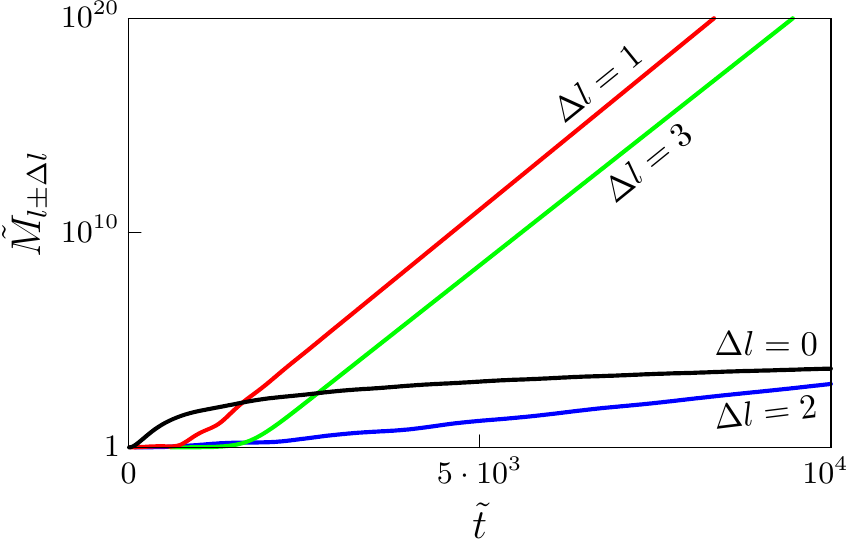}}
  \caption{Norms $M_{l\pm\Delta l}$ (logarithmic scale) of the linear
    perturbations evolving with time in the background of the
    $l=2$ Bose star at~$\lambda=0$.}
  \label{fig:instab2}
\end{figure}

To extract the exponentially growing instability modes
\begin{equation}
  \label{eq:4}
  \delta \psi ,\; \delta \bar{\psi} ,\; \delta \Phi \propto
  \mathrm{e}^{\mu t} \qquad \mbox{with} \qquad \mathrm{Re}\, \mu > 0\;,
\end{equation}
we evolve the axially--symmetric equations~(\ref{eq:64}) in real
  time $t$ using the numerical method of
Appendix~\ref{sec:axially-symm-simul}. The norms $M_{l\pm\Delta  l }(t)
\equiv m\int d^3 x |\delta \psi|^2$ of the perturbations
in the background of the $l=2$ star are shown in
Fig.~\ref{fig:instab2}. They indeed grow
exponentially\footnote{The 
  exponents of the graphs with $\Delta l=1$ and~$3$ are visibly close,
  though the one with $\Delta l=1$ grows faster. This near
  degeneracy is a peculiarity of the $l=2$ Bose star.},  
as predicted by our no--go theorem.  

In practice it is more convenient to keep the perturbations
finite. To this end we multiply $\delta \psi$, $\delta
\bar{\psi}$, and $\delta \Phi$ by a certain
complex factor $\Delta {\cal N}$ after every time step. The
  resulting renormalized solution approaches the profile of the 
fastest--growing instability  mode at large $t$, while the
  respective growth exponent equals $\mu = \Delta t^{-1}\ln \Delta
{\cal N}$. 

\begin{figure}
  \centerline{\includegraphics{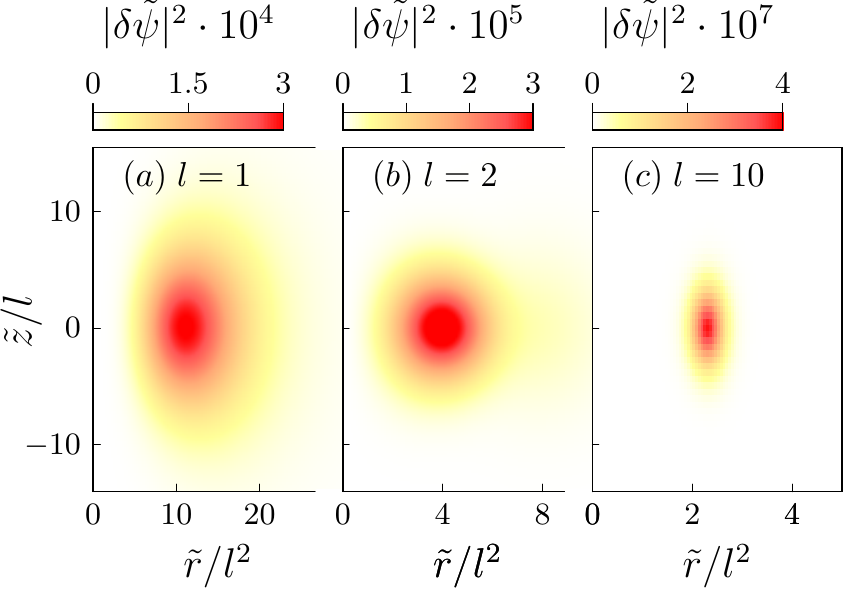}}
  \caption{Dominant instability modes $\delta \psi(r,\, z)$ of the rotating
    Bose stars at: (a)~$l=1$, $\Delta l=2$; (b)~$l=2$,
    $\Delta l=1$; and~(c)~${l=10}$, $\Delta l=6$. Model with
    $\lambda=0$ is considered.}
  \label{fig:modes}
\end{figure}

\begin{figure}
  \centerline{\includegraphics{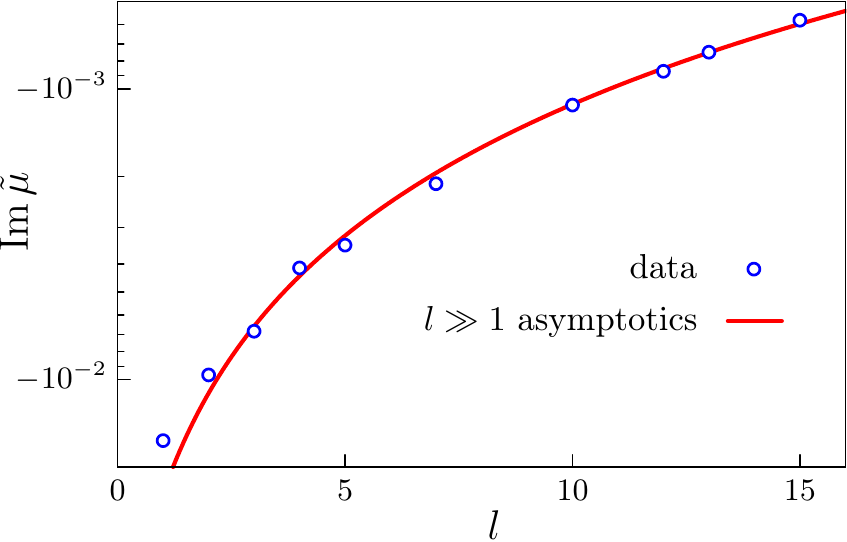}}
  \caption{Imaginary parts of the dominant instability exponents
    $\mathrm{Im}\, \mu$ computed in the backgrounds of
    rotating Bose stars with different $l$,
    cf.\ Fig.~\ref{fig:instab-exponent} and see Eq.~(\ref{eq:9}). We
    consider negligibly small boson self--coupling $\lambda=0$.
    Numerical data (circles) approach the analytic asymptotics
    (\ref{eq:5}) (lines) at large $l$.}
  \label{fig:muI}
\end{figure}

In Fig.~\ref{fig:modes} we demonstrate the dominant instability
modes of the Bose stars with $l=1$, 2, 
and~10. Like the background stars, they have toroidal
  forms. The exponents $\mu$ and 
angular momentum transfers $\Delta l$ of these 
  perturbations are listed in
Table~\ref{tab:parameters_instability} and shown in 
Figs.~\ref{fig:instab-exponent},~\ref{fig:muI}; see also 
Eq.~(\ref{eq:9}). These data reproduce the result of the
three--dimensional simulations at $l=1$ and approach analytic 
expressions of the next Section (solid line) at $l\gg 1$.

%%%%%%%%%%%%%%%%%%%%%%%%%%%%%%%%%%%%
\subsection{Self-interacting condensate}
\label{sec:self-inter-cond}
Using the procedure of Sec.~\ref{sec:no-self-interaction}, we compute
Bose stars at different nonzero $\lambda$. But this time we
restrict ourselves to the case $l=1$. Namely, restoring the last
term in Eq.~(\ref{eq:12}), we alternate Euclidean evolution steps
$\Delta \tau = i\Delta t$ with renormalizations of $\psi$ and SOR
sweeps for~$\Phi$ in Eq.~(\ref{eq:13}). This gives star
configurations with fixed~${\tilde{M}_s}$.

Notably, the iterations converge only at ${\lambda > \lambda_{cr}}$,
where the value of the critical coupling ${\lambda_{cr}<0}$  was already introduced in  
Eq.~(\ref{eq:25}). This means that  the fixed--mass stationary solutions  do
not exist at  couplings below $\lambda_{cr}$ (stronger attraction) or,
conversely, at a given negative $\lambda$ and overcritical mass $M_s
> M_{cr}^{(l=1)}$. Expressing the mass from
  Eq.~(\ref{eq:25}), one obtains, 
\begin{equation}
  \label{eq:29}
  M_{cr}^{(l=1)}   \approx (27.17 \pm 0.07)/(-G\lambda)^{1/2}
\end{equation}
at fixed $\lambda$.

The overcritical stars do not exist because as one can see numerically, at
fixed $\lambda<0$ the mass $M_s$ grows with  
$|\omega_s|$  until reaching the maximum ${M=M_{cr}^{(l=1)}}$ with
${dM_s/d\omega_s \approx 0}$ cf.~\cite{vk_71, Zakharov_2012}.  It is clear
that analytic continuation to larger $|\omega_s|$ at the other side of
the maximum would produce smaller--mass solutions rather than  the
heavy Bose stars.

The above critical behavior at $M>M_{cr}^{(l=1)}$ is the same as in
the case of non--rotating Bose
stars~\cite{Chavanis:2011zi}. Physically, it is caused by
self--attraction dominating in the dense Bose--Einstein
condensate and forcing the objects with large mass to collapse,
i.e.\ squeeze in a self--similar manner~\cite{Zakharov_2012,  
  Levkov:2016rkk}. The collapse ends with streams of relativistic
bosons leaving the condensate~\cite{Levkov:2016rkk}. Due to this
process, no stationary Bose stars with given mass exist 
at~${\lambda < \lambda_{cr}}$ or $M_s > M_{cr}^{(l=1)}$.

We obtain the value (\ref{eq:29}) of the critical mass by
  fitting the fixed--$\lambda$ numerical data for $ M_s(\omega_s)$
  with the parabola
\begin{equation}
  \label{eq:16}
  M_s(\omega_s) \approx M_{cr}^{(l=1)} - c_1 (\omega_s - \omega_{cr})^2
\end{equation}
in the region $\omega_s \approx \omega_{cr}$, where $M_{cr}^{(l=1)}$,
$c_1$, and $\omega_{cr}$ are the fit parameters. This produces
Eq.~(\ref{eq:29}) with errorbars estimating the sensitivity of the fit
to the lattice spacing and to the the interval of $\omega_s$. Inverting
Eq.~(\ref{eq:29}),  we get Eq.~(\ref{eq:25}).

Running the three--dimensional simulations of Sec.~\ref{sec:numer-illustr}
in the model with attractive self--interactions, we explicitly
verified that the $l=1$ Bose star indeed collapses at $\lambda <
\lambda_{cr}$, see the movie~[\onlinecite{movie}b].

In the opposite case ${\lambda > \lambda_{cr}}$, we study the
Bose star stability by adding self--interaction terms  to  
Eqs.~(\ref{eq:64}) and evolving the perturbations in real time. The
norms of $\delta \psi$, $\delta \bar{\psi}$, and $\delta \Phi$ grow 
exponentially at $\lambda < \lambda_0$, where
  $\lambda_0$ is previewed in Eq.~(\ref{eq:20}). Their exponents
$\mathrm{Re}\,\mu$ are shown in Fig.~\ref{fig:mu_lambda}. Thus, in the
entire region $\lambda_{cr}  < \lambda < \lambda_0$ the Bose star with
$l=1$ decays by shedding off its angular momentum.

The precise value~(\ref{eq:20}) of $\lambda_0$ is obtained by fitting
the numerical data for $\mathrm{Re}\, \mu (\lambda)$ with the
threshold function
\begin{equation}
  \label{eq:22}
  \mathrm{Re}\, \mu = \left[d_1 (\lambda_0 - \lambda)  + d_2
    (\lambda_0 - \lambda)^2\right]^{1/2}
\end{equation}
in the region near $\lambda_0$. Like before, the numerical errors are
estimated by varying the lattice spacing and interval of
$\lambda$.

At $\lambda> \lambda_0$, the perturbations remain bounded during
the entire real--time evolution. Fitting formally their norms
with the exponents, we obtain points in the right--hand side of
Fig.~\ref{fig:mu_lambda}. The respective values of
  $\mathrm{Re}\, \tilde{\mu}$ are all below
  $2\times 10^{-4}$. In fact, they are smaller than the expected numerical 
  precision and therefore consistent with ${\mathrm{Re}\, \mu \approx
    0}$, cf.\ Appendix~\ref{sec:axially-symm-simul}. Performing the
three--dimensional simulations, we checked that even strongly
perturbed star does not decay in this case, see the
movie~[\onlinecite{movie}c]. We conclude that the $l=1$ Bose stars are
absolutely stable at $\lambda > \lambda_0$ or
\begin{equation}
  \label{eq:28}
  M > M_{0}^{(l=1)} \approx 25.9/(G\lambda)^{1/2}\;,
\end{equation}
where Eq.~(\ref{eq:20}) was rewritten in terms of mass at a fixed coupling.

%%%%%%%%%%%%%%%%%%%%%%%%%%%%%%%%%%%%%%%%%%%%%
\section{Analytic solutions at $l\gg 1$}
\label{sec:analytic-solution-at}

%%%%%%%%%%%%%%%%%%%%%%%%%%%%%%%%%%%%%%%%%%%%%%
\subsection{Bose stars}
\label{sec:bose-stars}
At large $l$ the profiles of rotating Bose stars and their instability
modes can be evaluated analytically. Let us explain the idea of the
approximation using crude estimates. In this Section we
  consider the case of negligible self--coupling,~${\lambda=0}$.

\begin{figure}
  \centerline{\includegraphics{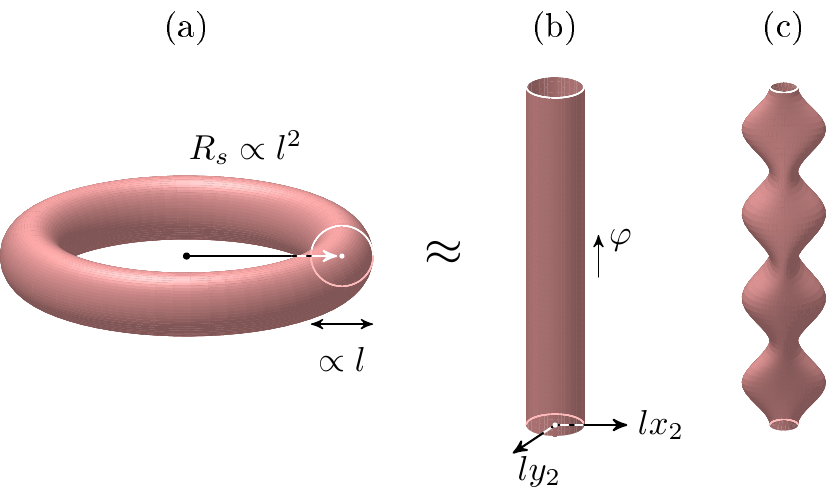}}
  \caption{(Not to scale) (a) Constant density surface of the
    rotating Bose star at large $l$. (b)~The same for the  
    equivalent object with cylindrical symmetry. (c)~Adding the
    perturbation.}
  \label{fig:torus} 
\end{figure}

The size $R_s$ of the fast--rotating Bose star is large. Indeed, it is
determined by balance between the gravitational and centrifugal forces,
$|\Phi_s| \sim GM_s/R_s \sim (l/mR_s)^2$, and 
therefore grows as
\begin{equation}
  \label{eq:8}
  R_s = 2\pi l^2 / (m^2 GM_s \alpha_l) \propto l^2
\end{equation}
at ${l\to   +\infty}$, where we introduced an order--one
parameter~$\alpha_l$. On the other hand, the typical momentum of
the condensed bosons is related to the depth of the star potential well:
${\Delta_{r,z} \psi_s   /\psi_s \sim m^2 \Phi_s \propto l^{-2}}$ and
hence $\partial_{r,\, z} \psi_s / \psi_s \propto l^{-1}$. This immediately
suggests that the large $l$ star has the form of a ring 
in Fig.~\ref{fig:torus}a with radius and thickness proportional 
to $l^2$  and $l$, respectively. 

The above property is explicit in numerical profiles obtained in the
previous Section. Indeed, the object with $l=10$ in
Fig.~\ref{fig:large-l-star} resembles the torus with two
essentially different radii. Naturally, we want to describe such
ring--like objects in coordinates~${\bm{x}_2 = (x_2,\, y_2)}$,
\begin{equation} 
  \label{eq:70}
  r = R_s + lx_2 \;, \qquad z = ly_2\;,
\end{equation}
which do not depend on $l$. Recall also that the size ${R_s\propto l^2}$
of the ring is controlled by the new parameter $\alpha_l$ that will be 
specified afterwards.

The above observation fixes the $l$ dependence of the binding energy
${\omega_s \equiv \omega_2/l^2}$ and of the fields,  
\begin{equation}
  \label{eq:67}
  \psi_s = l^{-2}\, \psi_2(\bm{x}_2) \,, \;\;\;\; \Phi_s =
  l^{-2}\Phi_2(\bm{x}_2) - \frac{l^2}{2m^2 R_s^2}\, ,
\end{equation}
where we again assumed that the Bose star mass~(\ref{eq:46}) does not
depend on $l$. Substituting the Ansatz~(\ref{eq:67}) into
Eqs.~(\ref{eq:13}), (\ref{eq:45}) and ignoring the terms suppressed by
$l^{-1}$, we arrive at equations for the ring profile,
\begin{align}
  \label{eq:68}
  &\omega_2 \psi_2 = -\frac{\Delta_2 \psi_2}{2m} + m\Phi_2 \psi_2\;,
  \\
  \label{eq:74}
  & \Delta_2 \Phi_2 = 4\pi mG |\psi_2|^2\;. 
\end{align}
Here and below $\Delta_2 \equiv \partial_{x_2}^2 + \partial_{y_2}^2$
is the two--dimensional Laplacian.

Apparently, Eqs.~(\ref{eq:68}), \eqref{eq:74} repeat the original
Schr\"odinger--Newton problem~(\ref{eq:13}),~(\ref{eq:45}), but in two 
dimensions. Thus, the section $\varphi  = \mbox{const}$ of our
large $l$ Bose star has the same profile as its nonrotating
low--dimensional brother. The extra  factor $\mathrm{e}^{il\varphi}$ in
Eq.~(\ref{eq:42}) ensures rotation.

It is clear now, why all rotating Bose stars are unstable at
large~$l$. The radii of their rings are so large that the respective
curvature effects do not even contribute into the leading--order 
equations~(\ref{eq:68}), (\ref{eq:74}). Hence, these stars are
equivalent to the cylindrical objects in Fig.~\ref{fig:torus}b, which
can be in turn deprived of the extra--dimensional momentum~$l$ by the
Galilean transformation. The resulting static configurations of the   
Bose--Einstein condensate are unstable with respect to decay into many
spherical drops with smaller surface tension, see
Fig.~\ref{fig:torus}c. 

It is natural to expect that the solution of Eqs.~(\ref{eq:68}),
(\ref{eq:74}) has circular symmetry in the $\bm{x}_2$ plane
i.e.\ depends on $r_2^2 
\equiv x_2^2 + y_2^2$. As a consequence, the original three--dimensional
star is also symmetric, with surfaces of constant density forming flat
toruses. We use this property to compute the star profile:  substitute
$\psi_2 = \psi_2(r_2)$ and  $\Phi_2 (r_2)$ into  Eqs.~(\ref{eq:68}),
(\ref{eq:74}) and solve the resulting ordinary differential
equations with the shooting method. This standard calculation is
summarized in Appendix~\ref{sec:two-dimensional-bose}. Notably, the
resulting function $\psi_2 (r_2)$ is real.

\begin{figure}
  \centerline{\includegraphics{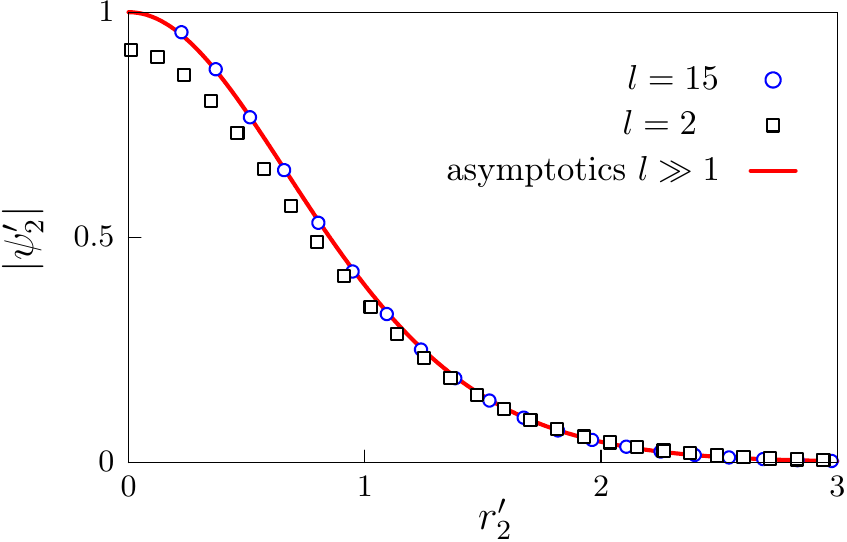}}
  \caption{Two--dimensional Bose star $|\psi_2(r_2)|$ (solid line)
    versus the sections of the rotating three--dimensional stars
    $|l^2\psi_s(R_s + lr_2,\, 0)|$ (points). In the latter
      case we determine the Bose star radius $R_s$ as  a position of
      the $|\psi_s|$ maximum at $z=\varphi = 0$, then use dimensionless units
      with parameter $v_0'$ obtained from Eq.~(\ref{eq:69}).
      The case~${\lambda=0}$ is considered.} 
  \label{fig:2d-profile}
\end{figure}
Numerically, it is again convenient to exploit dimensionless units
with $G = m = 1$ and restore physical terms afterwards. To this
end we rescale ${\bm{x}_2 = \bm{x}_2'/mv_0'}$ and ${\psi_2 =
  v_0'^2(m/G)^{1/2}   \psi_2'}$, $\Phi_2  = v_0'^2\, \Phi_2'$ like in 
three dimensions, but with the new parameter $v_0'$ selected to make 
$\psi_2'(0) = 1$. The two--dimensional profile $\psi_2'(r_2')$ is
demonstrated in Fig.~\ref{fig:2d-profile} (solid line). Notably, the
sections $\varphi = \mathrm{const}$ of the three--dimensional
rotating stars (points) approach this graph at~${l \to +\infty}$. 

Given $\psi_2$, we calculate the  Bose star parameters. Rewriting the
integral~(\ref{eq:46}) at large $R_s$ in two--dimensional
terms~\eqref{eq:70}, \eqref{eq:67} and performing the rescaling, we
arrive at the Bose star mass,
\begin{equation} 
  \label{eq:69}
  M_s = \frac{4\pi^2 (v_0')^2M_{2}'}{m^2 G^2 M_s \alpha_l} \;,
\end{equation}
where we used Eq.~(\ref{eq:8}) and computed the remaining
dimensionless integral ${M_2' \equiv \int d^2\bm{x}_2' \, |\psi_2'|^2 
  \approx 1.70}$. In practice one can use this relation to
  express the rescaling parameter $v_0'$ in terms of the total mass
  $M_s$.
  Similarly, the energy~(\ref{eq:56}) of 
the Bose star equals,  
\begin{equation}
  \label{eq:66}
  E_s = \frac{m^2 G^2 M_s^3}{8\pi^2 l^2} \; \alpha_l \left[\alpha_l +
    \frac12 + \ln(\beta \alpha_l/l^2) \right]\;.
\end{equation}
This time we extracted, in addition, $v_0'$ from Eq.~(\ref{eq:69}) and
introduced another numerical coefficient  $\beta \approx 2.86 \cdot
10^{-2}$, see Appendix~\ref{sec:two-dimensional-bose} for details.

We finally extremize the energy~(\ref{eq:66}) with respect to the
parameter $\alpha_l$ characterizing the Bose star radius $R_s\propto
l^2/\alpha_l$. This gives a nonlinear equation~(\ref{eq:71}) and
finishes construction of the large $l$ Bose star. Recall that
we already previewed the energy asymptotics~(\ref{eq:66}) in the
last element of Table~\ref{tab:bs_parameters} and in
Fig.~\ref{fig:energy_l}. The last graph roughly agrees with the
numerical data even at
$l\sim 1$ becoming more precise at larger~$l$. As always, 
the binding energy of the Bose particles inside the Bose star equals
$\omega_s = 3mE_s/M_s$, see Eq.~(\ref{eq:2}). 

%%%%%%%%%%%%%%%%%%%%%%%%%%%%%%%%%%%%%%%%%%%
\subsection{Instabilities}
\label{sec:instability-modes-1}
\label{eq:88}
Now, we evaluate exponentially growing modes destroying the
fast--rotating Bose stars.

To this end we rescale $l$ from the coordinates and background fields
in the linear equations (\ref{eq:64}) using
Eqs.~(\ref{eq:70}), (\ref{eq:67}). Then, substituting ${\delta \psi, \,
\delta \bar{\psi},\, \delta \Phi \propto \mathrm{exp}(\mu t)}$, we arrive at the
  leading--order eigenvalue problem
\begin{align}
  \notag
   - \mu_2 \, \eta & = \frac{p_\varphi^2 - \Delta_2}{2m}\; 
  \rho + m\psi_{2}\, \delta \Phi + (m\Phi_{2} - \omega_{2}) \,
  \rho,\\[.5ex]
  \label{eq:87}
  \mu_2\, \rho & = \frac{p_\varphi^2 - \Delta_2}{2m} \; 
  \eta+ (m\Phi_{2} - \omega_{2}) \, \eta\;,\\[.5ex]
  \notag
  \Delta_2 \delta \Phi & = p_\varphi^2\, \delta \Phi +  8\pi Gm \psi_2 \, \rho\;.
\end{align}
Here we recalled that $\psi_{2}(\bm{x}_2)$ is real and
  introduced ``real'' and ``imaginary'' perturbations 
  ${\rho(\bm{x}_2) \equiv (\delta \psi + \delta\bar{\psi})/2}$ and
  $\eta(\bm{x}_2)\equiv (\delta \psi -
  \delta\bar{\psi})/2i$. Besides, in Eqs.~(\ref{eq:87}) we traded
  the angular momentum transfer 
$\Delta l$ and complex exponent $\mu$ for the parameters
\begin{equation}
  \label{eq:85}
  p_\varphi = l\Delta l/R_s\;, \qquad \mu_2 = l^2 (\mu + ip_\varphi/mR_s)\;.
\end{equation}
Solving the eigenvalue problem (\ref{eq:87}), one can find all
vibrational modes of the Bose star at a given $\Delta l$ and determine
their exponents $\mu$. Notably, the same problem with $p_\varphi=0$
describes vibrations of the two--dimensional star $\psi_2(r_2)$ which
  is stable. We will therefore focus on the instability modes with
${\mathrm{Re}\,\mu > 0}$ at ${p_\varphi\ne 0}$.

Notably, these exponentially growing perturbations with
${\mathrm{Re}\,\mu > 0}$ have several properties proved in 
Appendix~\ref{sec:two-dimensional-bose}. First, their eigenvalues
$\mu_2$ and profiles $\rho$, $\eta$, $\delta \Phi$ are 
real. Expectedly so, since Eqs.~(\ref{eq:87}) are
real--valued. As a consequence, $\mu_2$ and $p_\varphi$
parameterize real and imaginary parts of the original exponent 
$\mu$ via Eqs.~(\ref{eq:85}). Second, all instability modes vanish
at infinity,
\begin{equation}
  \label{eq:89}
  \rho ,\; \eta,\; \delta \Phi \to 0 \;\;\;\; \mbox{as} \;\;\;\; |\bm{x}_2|
  \to +\infty\;.
\end{equation}
Indeed, less localized solutions of Eqs.~(\ref{eq:87}) describe waves
oscillating with real frequencies $i\mu_2$ in the far--away region 
with $\psi_2 \approx 0$. Third and finally, the instability modes are 
rotationally--symmetric from the two--dimensional viewpoint,
i.e.\ depend only on $r_2 = |\bm{x}_2|$. This last fact agrees with
the intuitive figure~\ref{fig:torus}c.

We explicitly compute the profiles of the exponentially growing
perturbations using the same strategy as before. Substitute the
rotationally invariant Ansatz $\rho(r_2)$, $\eta(r_2)$, $\delta
\Phi(r_2)$ into Eqs.~(\ref{eq:87}) and perform rescaling with the 
parameter $v_0'$, e.g.\ $\rho = v_0'^2 (m/G)^{1/2}\rho'(r_2')$. This
gives the system of ordinary differential equations with two
dimensionless constants: eigenvalue ${\mu_2' =\mu_2/(mv_0'^2)}$ and
rescaled extra-dimensional momentum ${p_{\varphi}' =
  p_\varphi/(mv_0')}$. After that apply the shooting method to
solve the equations with regularity conditions at the origin and
falloff conditions~(\ref{eq:89}) at infinity, see
Appendix~\ref{sec:two-dimensional-bose} for details. 

\begin{figure}[htb]
  \centerline{\includegraphics{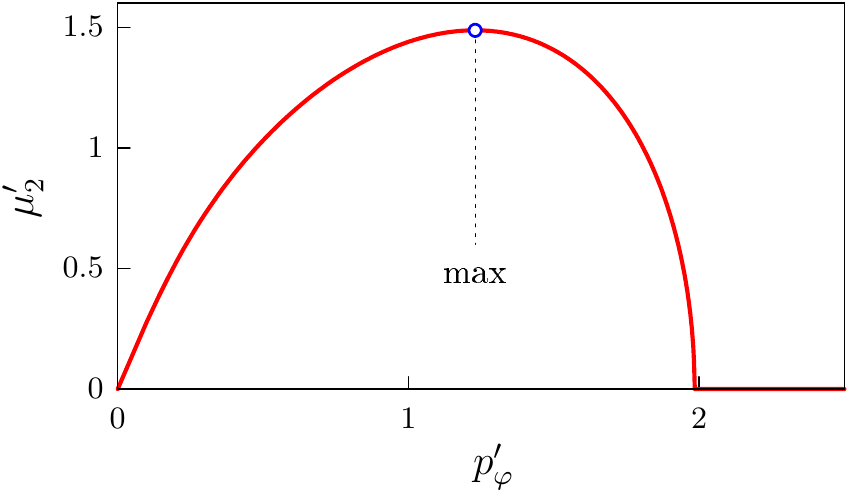}}
    \caption{Eigenvalue $\mu_2$ of the instability mode
    as a function of the extra--dimensional momentum
    $p_\varphi$.}
  \label{fig:pz2graph}
\end{figure}

We find precisely one instability mode at $0 < p_\varphi' < 2$,
and no modes outside of this interval. The respective
eigenvalue $\mu_2 (p_\varphi)$ is plotted in Fig.~\ref{fig:pz2graph}.
Since by itself the two--dimensional Bose star is
stable, there are no instabilities at $p_\varphi =0$. In the
opposite limit of large $p_\varphi$ rotational energy makes the
operators in the right--hand sides of Eqs.~(\ref{eq:87})
positive--definite and drives $\mu_2^2$ to negative values.

 The fastest--growing perturbation corresponds to the maximum of the
 graph~\ref{fig:pz2graph} at $\mu_2' \approx 1.49$ and ${p_\varphi'
   \approx 1.23}$ (point). Rescaling back to physical units and
 using Eqs.~(\ref{eq:85}),~(\ref{eq:8}), we obtain real and imaginary
 parts of the  growth exponent $\mu$ in Eq.~(\ref{eq:9}) with coefficients 
 \begin{align}
   \label{eq:73}
   & \mathrm{Re}\, \tilde{\mu} = \frac{\mu_2'\, \alpha_l}{(2\pi l)^2
     M_2' } \approx 2.22 \cdot 10^{-2} \; \frac{\alpha_l}{l^2}\;, \\
   \label{eq:63}
   & \mathrm{Im}\,\tilde{\mu} = - \frac{p_\varphi' \;\alpha_l^{3/2}}{(2\pi l)^2
     (M_2')^{1/2} } \approx - 2.39 \cdot  10^{-2} \; \frac{\alpha_l^{3/2}}{l^2}\;.
 \end{align}
 Besides, the first of Eqs.~(\ref{eq:85}) fixes the angular momentum
 transfer driving the instability,
 \begin{equation}
   \label{eq:72}
   \Delta l  =\left[ \frac{ l \, p_\varphi'}{\alpha_l^{1/2} (M_2')^{1/2}}\right]
     \approx \left[ \frac{0.944 \cdot l}{\alpha_l^{1/2}} \right] \;,
 \end{equation}
where $[\cdot ]$ denotes the closest integer. Recall that
$\alpha_l$ satisfy Eq.~(\ref{eq:71}). We previewed the above
asymptotic expressions in Eq.~(\ref{eq:5}) of the Introduction and
visualized them in Figs.~\ref{fig:instab-exponent}, \ref{fig:muI}. Let 
us repeat that the numerical results of
Sec.~\ref{sec:instability-modes} approach the asymptotic expressions 
at large $l$ in all figures. 

We finish this Section with a forecast on the number of non--spinning
Bose stars that can form in the decay of the star with large $l$. The
mass density of the fastest--growing instability mode is proportional to
$\cos(\varphi\Delta l +  \mathrm{const})$ and therefore has $\Delta l$
maxima along the ring,  
see Eqs.~(\ref{eq:3}). Thus, $\Delta l$ non--spinning objects with 
mass $ M_s / \Delta l $ appear at the first stage of the process,
moving in a carousel around the common center.
It is instructive to compare their net energy with that of the
original Bose star,
\begin{equation}
  \label{eq:92}
  \frac{E_{s,\, l}(M_s)}{\Delta l \; E_{s,\, 0}(M_s/\Delta l)} =
  \frac{\alpha_l (\alpha_l+1)\Delta l^2}{8\pi^2 |\tilde{E}_{s,\, 0}|
    l^2} \approx 0.21 (\alpha_l + 1)\,,
\end{equation}
where the indices and arguments of $E_s$ indicate the mass and angular
momentum of the respective Bose star; we used Eqs.~(\ref{eq:21}),
(\ref{eq:66}), (\ref{eq:72}) and read off $\tilde{E}_{s,\, 0}$ from
Table~\ref{tab:bs_parameters}. Equation (\ref{eq:92}) implies that the
Bose stars with $l \geq 32$ have lower total energy than the chain of
$\Delta l$ non--spinning smaller--mass stars. Thus, the latter objects
should remain bound together for some time, like the two--star state
in  Fig.~\ref{fig:inst-comic}, until~--- possibly~--- subdominant
instabilities will destroy the chain. On the other hand,
some rotating stars with $l<32$ may directly decay into $\Delta
l$  non--spinning ones.

%%%%%%%%%%%%%%%%%%%%%%%%%%%%%%%%%%%%%%%%%
\section{Discussion}
\label{sec:discussion}
In this paper we analytically proved that rotating nonrelativistic
Bose stars are unstable at any angular momentum if self--coupling of
their bosons is attractive or negligible, $\lambda  \leq 0$. This
result is relevant for the  popular models with QCD axionic or fuzzy dark
matter. We also demonstrated that in models with repulsive
self--interactions ($\lambda>0$) the $l=1$ star is unstable at 
masses below ${M_{s,\, 0} \approx  25.9/(\lambda G)^{1/2}}$ and
absolutely stable at ${M_s > M_{s,\,0}}$, cf.\ Eq.~\eqref{eq:20}. 

We computed the lifetimes of the unstable rotating stars in
Eqs.~\eqref{eq:9}~---~\eqref{eq:15} and in
Table~\ref{tab:parameters_instability}. They are always
comparable to the inverse binding energies $\omega_s^{-1}$ of the
Bose stars and  smaller than the free--fall times in their gravitational
fields. Thus, the rotating stars~(\ref{eq:42}) cannot nucleate in
realistic formation scenarios~\cite{Seidel:1993zk,   Schive:2014dra,
  Levkov:2018kau,   Eggemeier:2019jsu, Eggemeier:2019jsu,
  Chen:2020cef} and in fact, cannot be even considered as
long--living quasi--stationary states. This observation has a
number of phenomenological consequences.

First, the scenario~\cite{Hertzberg:2018zte} with rotating axion
stars reaching threshold for the explosive parametric
radioemission~\cite{Tkachev:1986tr,Levkov:2020txo,Hertzberg:2020dbk,
  Amin:2020vja, Amin:2021tnq}
cannot be realized. One still can consider emission during the
intermediate stages when dense and short--living rotating configuration
shakes off its angular momentum. But a specific formation scenario for
the latter should be suggested in the first place.

Second, instability of rotating Bose stars provides a universal
mechanism to destroy the angular momentum. One can imagine e.g.\ 
that a subset of dark matter Bose stars collapses gravitationally
into black holes with suppressed spins. This is possible in models
with positive self--coupling or in axionic models with near--Planckian
decay constants~\cite{Helfer:2016ljl}. Formation of such non--spinning
black holes may explain observational hints 
in~\cite{LIGOScientific:2018mvr, LIGOScientific:2018jsj, Abbott:2020jks}.

In this paper we also developed an analytic description of fast--rotating
Bose stars with ${l\gg 1}$. This technique is complementary to the 
numerical methods at moderately small $l$, it gives parameters
and lifetimes of stars as systematic expansions in~$l^{-1}$,
cf.\ Eqs.~(\ref{eq:5}), \eqref{eq:66}. We saw that such analytics is
crudely applicable even at $l\sim 1$, and it becomes precise at
higher $l$, see Figs.~\ref{fig:instab-exponent},~\ref{fig:energy_l}.  

Our analytic method is based on a simple observation that the large $l$
Bose stars have forms of parametrically thin rings,
cf.\ Figs.~\ref{fig:large-l-star} and~\ref{fig:torus}a. As a
consequence, their sections and instability modes can be obtained by   
solving certain ordinary differential equations. This approach can be
easily generalized to nontrivial situations: the case of nonzero
self--coupling $\lambda \ne 0$, relativistic model with complex field
as in Refs.~\cite{Sanchis-Gual:2019ljs, DiGiovanni:2020ror,
  Siemonsen:2020hcg}, rotating star in the external gravitational
potential, etc.

Although the $l=1$ Bose star becomes stable at sufficiently strong
repulsive self--couplings $\lambda > \lambda_0$,
Eq.~(\ref{eq:20}), the fate of the higher $l$ objects is far less
trivial. In models with dominating   
self--repulsion the $l\geq 2$ vortices decay~\cite{LL9,
    Nugaev:2014iva} into elementary
ones with $l=1$, and the latter spread uniformly over the available
volume. This suggests that the axially--symmetric $l\geq 2$
configurations~(\ref{eq:42}) are unstable at any~$\lambda$,
and the real question is whether they decay into gravitationally
bound objects with~$l$ elementary vortices inside, or most of the
vortices
migrate to the periphery of the system and disappear in the 
debris. While this paper was approaching completion, a first study of
such process has appeared~\cite{Nikolaieva:2021owc}. 

Finally, let us remark that although formation of the rotating Bose
stars requires fine--tuning of the initial data or a special mechanism,
decays of these objects are so complex and aesthetically pleasing that
their studies may have a scientific value of their own. Indeed, we
expect higher $l$ stars to break into $\Delta l \propto l$
non--spinning  components oscillating and orbiting around the
mutual center, cf.\ the movie~[\onlinecite{movie}a]. This state should exist
for some time  until possibly breaking due to subdominant
instabilities. After that some components may tidally disrupt, and the
others survive. In the case 
of attractive self--interactions the component objects may appear
overcritical and collapse as bosenova bursting into relativistic
axions~\cite{Levkov:2016rkk} or photons~\cite{Levkov:2020txo},
since they are not protected by the centrifugal barriers anymore.

%%%%%%%%%%%%%%%%%%%%%%%%%%%%%%%%%%%%%%%%%%%

\acknowledgments

Instabilities of rotating Bose stars were studied within the
framework of the RSF grant 16-12-10494. The rest of this paper was 
funded by the Foundation for the Advancement of Theoretical Physics 
and Mathematics 
``BASIS.'' Numerical calculations were performed on the Computational
cluster of the Theory Division of INR~RAS.

%%%%%%%%%%%%%%%%%%%%%%%%%%%%%%%%%%%%%

\appendix

%%%%%%%%%%%%%%%%%%%%%%%%%%%%%%%%%%%%%
\section{Three-dimensional simulation}
\label{sec:numerical-methods}
In Sec.~\ref{sec:numer-illustr} we discuss simulations in 
three--dimensional box ${-L/2 < x,\,   y,\, z \leq L/2}$ with 
periodic  $\psi$ and $\Phi$. Consistency
requires modification of  Eq.~(\ref{eq:13}) to
\begin{equation} 
  \label{eq:11}
  \Delta \Phi = 4 \pi G (m|\psi|^2 - M/L^3)\;,
\end{equation}
where the new term with total mass $M$ vanishes as $O(L^{-3})$ in the
infinite--volume limit. We discretize $x$, $y$, and $z$ with uniform
lattice steps $\delta = L/N$ and place the fields ${\psi = \psi_{n,\, m,\,
    k}}$, $\Phi_{n,\,   m,\, k}$ on  the lattice sites
$\{x_n, \,  y_m,\, z_k\} \equiv \{n\delta,\,  m\delta,\,  k
\delta\}$. We perform all calculations in the dimensionless  units of
Sec.~\ref{sec:rotating-bose-stars}. 

Importantly, our cubic lattice is invariant with respect to $\pi/2$
rotations $\hat{R}_{\pi/2}$. Since $\hat{R}_{\pi/2}^4 = 1$, the latter
have four eigenvalues $\pm i$ and $\pm  1$, and the respective
eigenfunctions $\psi_{l}$ satisfy  Eq.~(\ref{eq:59}) with $0 \leq l
\leq 3$. In Eq.~(\ref{eq:60}) we  decompose $\delta\psi \equiv \psi -
\psi_s\,  \mathrm{e}^{-i\omega_s   t}$ into the sum of eigenfunctions $\psi_l$
using the projectors $\hat{\Pi}_{l}$,
\begin{multline}
  \label{eq:62}
  \psi_{l;\, n,\, m,\, k} = \hat{\Pi}_{l} \delta\psi_{n,\, m,\, k} =
  \frac14 \big[\delta \psi_{n,\, m,\, k} + 
    \mathrm{e}^{i\pi l/2} \delta \psi_{m,\, -n,\, k} \\
    +  \mathrm{e}^{i\pi l} \delta \psi_{-n,\, -m,\, k} + \mathrm{e}^{3i\pi
      l/2} \delta\psi_{-m,\,  n,\, k}\Big]\;.
\end{multline}
The operator $\hat{\Pi}_1$ with $l=1$ is used in the numerical
procedure below.

We evolve the Gross--Pitaevskii--Poisson equations (\ref{eq:12}),
(\ref{eq:13}) using fourth--order pseudospectral
method~\cite{Yoshida:1990zz, Levkov:2018kau}. In a nutshell, this
accounts to performing Fast Fourier Transforms at every time step:
first to advance the wave function ${\psi(t +   \Delta t) =
  \mathrm{e}^{-i \hat{H} \Delta t} \;\psi(t)}$ with precision
$O(\Delta t^5)$, then to solve the Poisson equation for the
gravitational potential; here $\hat{H}$ is the operator in the
right--hand side of Eq.~(\ref{eq:12}). Notably, this numerical scheme
equally treats all spatial coordinates and therefore commutes
with the $\pi/2$ rotations.

The same evolution in Euclidean time ${\tau = it}$ multiplies $\psi$
with $\mathrm{e}^{-\hat{H} \Delta \tau}$ and therefore minimizes the
energy of the configuration. Specifically, in
Sec.~\ref{sec:numer-illustr} we obtain the $l=1$ Bose star by
iterations ${\psi \to \Delta {\cal N}\mathrm{e}^{-\hat{H}\Delta
    \tau}\, \hat{\Pi}_{z}\, \hat{\Pi}_{1} 
  \, \psi}$, solving the Poisson equation at every step. Here the
operator $\hat{\Pi}_1$ projects $\psi$ onto the eigensector of the
$\pi/2$ rotations with the eigenvalue  $\mathrm{e}^{i\pi/2}$, another
projector ${\hat{\Pi}_z    \psi_{n,\, m,\,  k} \equiv \frac12(\psi_{n,\, m,\, 
    k} + \psi_{n,\, m,\,  -k})}$ stabilizes the configuration in the
$z$ direction by imposing $z\to -z$ symmetry, Euclidean propagator
$\mathrm{e}^{-\hat{H}\Delta \tau}$ kills the  higher--energy
components of $\psi$, whereas the normalization  factor $\Delta {\cal N}$
fixes the total mass $\int d^3 \tilde{x} \, |\tilde{\psi}|^2 = 1$ in
the rescaled units. At the end  of the relaxation the value of $\psi$
equals $\psi_s$ and the normalization factor $\Delta {\cal N} = 
\mathrm{exp}(\tilde{\omega}_s \Delta \tilde{\tau})$ gives the binding
energy $\tilde{\omega}_s$. 

\begin{figure}
  \centerline{\includegraphics{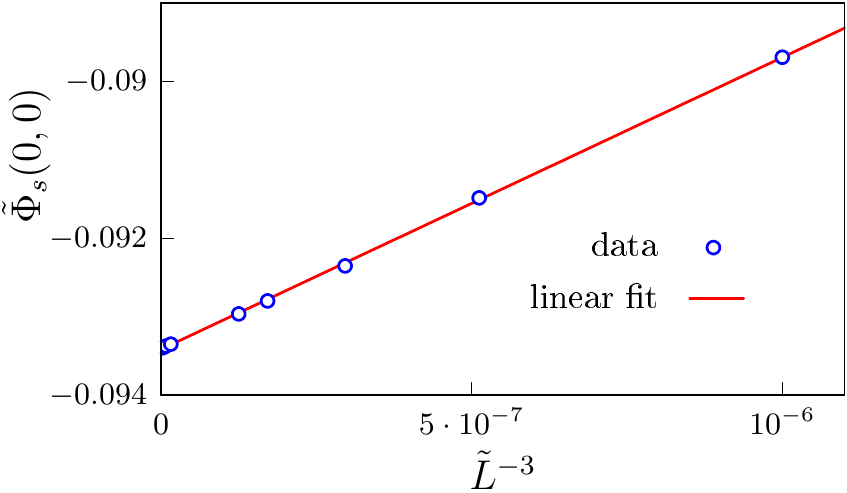}}
  \caption{The potential in the center $r = z = 0$ of the $l=1$
    Bose star as a function of the box size $L$ after the shift by
    $\Phi_0$; the case with no self--interactions is considered.}
  \label{fig:UL}
\end{figure}

In the periodic box, it is convenient to fix the constant part of the Bose
star potential using condition ${\int d^3
  \bm{x} \, \Phi_s = 0}$. We 
restore the standard terms by shifting ${\Phi_s \to \Phi_s + \Phi_0}$
and  $\omega_s \to \omega_s +\Phi_0$, where the constant $\Phi_0$ 
ensures Virial relation, e.g.\ $E_s = \omega_s N_s/3$ at
$\lambda=0$. After the shift, the finite--size effects in
the energetic
quantities reduce to $O(L^{-3})$, 
see  Fig.~\ref{fig:UL}. We do not  perform this shift at $\lambda \ne
0$ because it does not affect the data in Fig.~\ref{fig:mu_lambda}. 

In practice we exploit $128^3$ lattice in the $\tilde{L}=200$ box 
and use the time steps $\Delta \tilde{t} = \Delta
\tilde{\tau} = 0.5$. After  $10^4$ Euclidean iterations the
configuration with $l=1$ stabilizes at the relative level $\delta
\psi_s /\psi_s\sim 10^{-15}$ which is comparable to the round--off
errors. The relative effects of the finite lattice spacing $\delta$ and
discrete time step are of order $2\cdot 10^{-14}$ and $2\cdot 
10^{-12}$, respectively; they are estimated by using $256^3$ lattice
and step $\Delta \tilde{\tau} = 0.25$. The 
largest numerical artifacts come from  the finite--volume
cutoff: increasing $\tilde{L}$ by a factor of two, we obtain $\psi_s$ 
with relative corrections of order $10^{-4}$.

Next, we add the perturbation~(\ref{eq:58}) to the Bose star and
evolve the resulting configuration in real time to $\tilde{t} =
10^4$. The total energy and mass of the solution are stable
during the entire evolution up to relative corrections of order $2\cdot
10^{-9}$ and  $5\cdot 10^{-11}$, respectively. At the same time, 
numerical inaccuracies in $\psi$ grow exponentially because the
evolution is unstable. Nevertheless, $\delta$ and $\Delta t$
discretization errors always stay below $\delta \psi/\psi  
< 10^{-7}$ and $10^{-5}$, while relative finite--volume errors remain
smaller than $10^{-2}$ at $\tilde{t} \lesssim 9500$ reaching 
10\% level only at the very end of the simulation.

%%%%%%%%%%%%%%%%%%%%%%%%%%%%%%%%%%%%%
\section{Axially--symmetric code}
\label{sec:axially-symm-simul}
To compute the ${l\geq 2}$ stars numerically, we introduce ${N_r \times
  N_z}$ lattice with uniform spacings $\delta_r$, $\delta_z$ in  
axial coordinates $r$, $z$. The sites of this lattice ${(r_j,\,
  z_k)\equiv   (j\delta_r,\, k\delta_z)}$ fill a large cylindrical  
region ${0\leq   r_j \leq L_r}$ and ${0\leq z_k   \leq   L_z}$ in the 
upper half of the three--dimensional  space. We store the
field values ${\psi_{j,\, k} \equiv \psi(r_j,\, z_k)}$ and $\Phi_{j,k}$ on
the lattice sites and reconstruct them at $z <  0$ using the
symmetry ${\psi(r,\, -z) = \psi(r,\, z)}$, ${\Phi(r,\,
-z) = \Phi(r,\, z)}$. We use dimensionless units  with $\tilde{m} = 
\tilde{G} = \tilde{M}_s = 1$ introduced in
Sec.~\ref{sec:rotating-bose-stars}.

The Laplacians in Eqs.~(\ref{eq:12}), (\ref{eq:13}) are discretized in
the standard second--order manner:
\begin{multline}
  \label{eq:19}
  \Delta \psi_{j,\, k} = (\psi_{j,\, k+1} + \psi_{j,\, k-1} -
  2\psi_{j,\, k})/\delta_z^2  \\ + (\psi_{j+1,\, k} + \psi_{j-1,\, k} -
  2\psi_{j,\, k})/\delta_r^2\\
  + (\psi_{j+1,\, k} -  \psi_{j-1,\, k})/(2\delta_r
  r_j)  - l^2 \psi_{j,\, k}/r_j^2\;, 
\end{multline}
where similar expression for $\Delta \Phi_{j,k}$ has no last
term. We supply the lattice equations with the regularity conditions at
the symmetry axis\footnote{At $l=0$ we use $\psi_{-1,\, k}
  = \psi_{1,\, k}$.} $r= 0$: $\psi  = \partial_r \Phi =0$ or
\begin{equation}
  \label{eq:23}
  \psi_{0,\, k} = 0\;, \qquad \qquad \Phi_{-1,\, k} = \Phi_{1,\, k}\;.
\end{equation}
Boundary conditions at $z = z_0=0$ follow from the ${z\to -z}$
reflection symmetry: 
\begin{equation}
  \label{eq:24}
  \psi_{j,\, -1} = \psi_{j,\, 1}\;, \qquad \Phi_{j,\, -1} = \Phi_{j,\,
    1}\;.
\end{equation}
Finally, we impose relevant falloff conditions at the ``infinite''
lattice boundaries $r= L_r$ and $z=L_z$. In there, the wave function 
vanishes,
\begin{equation}
  \label{eq:26}
  \psi_{j,\, k} = 0  \qquad \mbox{at} \qquad  \mbox{$j=N_r - 1$ or $k =
    N_z - 1$}\;,
\end{equation}
and the potential is close to the the asymptotics ${\Phi \approx -
  GM/(r^2 + z^2)^{1/2}}$. The latter condition can be written in the
mass--independent form:
\begin{equation}
  \label{eq:14}
  \Phi_{N_r - 1,\, k} = \Phi_{N_r - 2,\, k}\left(\frac{r_{N_r - 2}^2 +
    z_k^2}{r_{N_r - 1}^2 + z_k^2}\right)^{1/2}\;,
\end{equation}
and similarly at $z = L_z$ and arbitrary $r_j$. To sum up, the above
discretization gives a set of evolution and Poisson equations at the
internal lattice sites with boundary values of the fields fixed by
Eqs.~(\ref{eq:23})~---~(\ref{eq:14}). 

We solve Eq.~(\ref{eq:13}) for $\Phi_{j,\, k}$ with the standard
red--black SOR method~\cite{NR}. After every relaxation sweep we
evolve\footnote{Since the Euler formula~\eqref{eq:18} is unstable, we
  upgrade it to a semi--implicit method: replace $\psi_{j,\, k}^{(n)} \to
  \psi_{j,\, k}^{(n+1)}$ in all diagonal terms of the operator $\hat{H}$
  and express $\psi^{(n+1)}_{j,\, k}$ from the resulting equation.}
the wave  function in Euclidean time by~${\Delta \tau  = i\Delta t}$,   
\begin{equation}
  \label{eq:18}
  \psi_{j,k}^{(n+1)} = \psi_{j,k}^{(n)} - \Delta \tau \, \hat{H} \psi^{(n)}_{j,k}\;,
\end{equation}
where $n$ indexes the sweeps and $\hat{H}$ denotes the discretized
operator in the right--hand side of Eq.~(\ref{eq:12}). Like before,
the evolution \eqref{eq:18} kills all excited energy levels in $\psi$
at a given $l$. Finally, we rescale $\psi_{j,k} \to
\Delta {\cal N} \, \psi_{j,k}$ to keep the total mass $\tilde{M}_s
  = 1$ fixed and then proceed to the next relaxation sweep. We
  decrease the time steps from $\Delta \tau  \propto \delta_{r,\,
    z}$ in the beginning of the relaxation to $\propto \delta_{r,\, z}^2$
  at the end of it.

The iterations converge producing the Bose stars
up to corrections $O(\delta_{r,z}^2)$  in lattice steps
and\footnote{Because Eq.~(\ref{eq:14}) ignores the dipole part in the
  Bose star gravitational potential.} $O(L_{r,z}^{-3})$ in box
size. Changing the parameters, we numerically confirmed the scalings of
the numerical errors with $\delta_{r,z}$ and $L_{r,z}$. The energy of
the Bose star is given by the discretized integral~(\ref{eq:56}),
while $\tilde{\omega}_s = 3\tilde{E}_s$ at $\lambda=0$, see 
Eq.~(\ref{eq:2}).  

In practical  computations we use lattices ranging between $N_r\times
N_z = {101\times 101}$ and $1501\times 1501$. We enlarge them
by a factor of two to control the discretization errors which never
exceed $\delta \psi_s / \psi_s < 10^{-2}$. Our box sizes $L_{r,\, z}
\propto l^2$ strongly vary with $l$ to encompass the Bose 
stars: from ${\tilde{L}_r = \tilde{L}_z = 100}$ at $l=1$ to
$22500$  at $l=15$. This keeps the relative finite--volume inaccuracies
below~$10^{-6}$. Finally, we ascertain that
the axially--symmetric profiles of the $l=0,\, 1$ Bose stars coincide with
the ones from the  three--dimensional code within the
expected $1\%$ accuracy.

Once the Bose star is obtained, we evolve Eqs.~(\ref{eq:64}) in real
time thus extracting the fastest--growing linear instability
mode. We use the same second--order discretization as before and
similar boundary conditions to
Eqs.~(\ref{eq:19})~---~(\ref{eq:26}). Since the real--time
evolution is more demanding to computational resources, we exploit smaller
lattices in smaller boxes at the cost of lower precision. Now,
$N_r\times N_z$ range between $101 \times 101$ and $1001 \times 1001$,
whereas the box sizes vary within the interval $10^2 \leq \tilde{L}_r
,\, \tilde{L}_z\leq 10^3$. Our time evolution uses Crank–Nicolson
steps~\cite{NR} with $\Delta \tilde{t} = 
0.2\, 
\tilde{\delta}_{r,\, z}^2/l'^2$. After every step we perform one SOR sweep for 
the $\delta \Phi$ equation. Then we multiply\footnote{This
  renormalization is switched off in Fig.~\ref{fig:modes}.} $\delta \psi$, $\delta 
\bar{\psi}$ and $\delta \Phi$ by a constant $\Delta {\cal N}$, and
proceed to the next time step. We stop the procedure when the
rescaled perturbations stabilize at the relative
level~$10^{-13}$. At the final step we compute the complex exponent of
the perturbation: $\tilde{\mu} = \Delta \tilde{t}^{-1} \ln \Delta 
{\cal N}$. Changing $l'$, we select the dominant
mode with maximal~$\mathrm{Re}\, \tilde{\mu}$. 

Like before, we estimate numerical precision by varying $N_{r,\,
  z}$, $L_{r,\,z}$, and $\Delta t$. All relative inaccuracies stay below 1\%,
although this time the largest errors are related to the comparable
finite--volume and discretization effects.

%%%%%%%%%%%%%%%%%%%%%%%%%%%%%%%%%%%%%
\section{Two--dimensional Bose stars}
\label{sec:two-dimensional-bose}
In this Appendix we numerically solve Eqs.~(\ref{eq:68}),
(\ref{eq:74}) for the profile of the two--dimensional Bose
star. To this end we absorb $\omega_2$ into the potential, 
\begin{equation}
  \label{eq:75}
  U_2(\bm{x}_2) \equiv \Phi_2(\bm{x_2}) -  \omega_2/m\;,
\end{equation}
introduce radial coordinate $r_2 = |\bm{x}_2|$ and perform coordinate
and field rescalings with parameter ${(v_0')^2 =
  (G/m)^{1/2}\, \psi_2(0)}$, as described in
Sec.~\ref{sec:bose-stars}. Our choice of $v_0'$ fixes the central
value of the dimensionless field to ${\psi_2'(0) = 1}$.

Assuming rotational symmetry, we obtain ordinary differential
equations for $\psi_2'(r_2')$ and ${U_2'(r_2')}$,
\begin{subequations}
  \label{eq:35}
\begin{align}
  \label{eq:78}
  & \partial_{r_2'} ( r_2' \partial_{r_2'} \psi_2' ) = 2 r_2' U_2'
  \psi_2'\;, \\
  \label{eq:79}
  & \partial_{r_2'} ( r_2' \partial_{r_2'} U_2') = 4\pi r_2' |\psi_2'|^2\;.
\end{align}
\end{subequations}
We supply them with the regularity conditions at the origin,
\begin{equation}
  \notag
  \partial_{r_2'} U_2'(0) = \partial_{r_2'} \psi_2' (0) = 0\;, \;\;
  \psi_2'(0) = 1 \;, \;\; U_2'(0) = U_0'\;,
\end{equation}
where $\psi_2'(0)=1$ follows from the field rescaling and we
introduced new parameter $U_0'$. 

We numerically solve Eqs.~(\ref{eq:35}) from the above initial data
to large $r_2'$, and then tune $U_0'$ to ensure falloff of the wave function
at infinity: $\psi_2' \to 0$ as ${r_2'   \to   +\infty}$. This gives
${U_0' \approx -2.07}$ and a configuration in
Fig.~\ref{fig:2d-profile} (solid line). After that we compute the
  dimensionless integral
\begin{equation}
  \label{eq:36}
  M_2' \equiv \int d^2 \bm{x}_2' \,  |\psi_2'|  \approx 1.7
\end{equation}
in the expression (\ref{eq:69}) for the Bose star mass.

Let us now apply the above object to find the profile of the fast--rotating
three--dimensional Bose star. In two dimensions, the  gravitational
potential logarithmically grows at infinity, 
\begin{equation}
  \label{eq:80}
  U_2' \to M_2' \, \ln \left(r_2'^2M_2'/64\beta\right)\;, \;\; \mbox{as} \;\;
  r_2' \to +\infty\;,
\end{equation}
where the numerical constant ${\beta \approx 2.86 \cdot 10^{-2}}$
parameterizes its constant part. As a consequence, 
we cannot separate $U_2$  into $\Phi_2$ and $\omega_2$ using
two--dimensional logic alone.

To do that, we compute the original three--dimensional potential at the
spatial point $z=0$ and ${r = R_s + lr_2}$ located 
parametrically far away from the ring core, ${r_2 \gg O(1)}$, and yet
belonging to the applicability region ${r_2 \ll R_s/l}$ of
Eqs.~(\ref{eq:35}). In this case the details of the ring 
profile are indiscernible, and we write
\begin{align}
  \notag
  \Phi_s &\approx - \frac{GM_s}{2\pi} \, \oint
  \frac{d\varphi}{\sqrt{4 (R_s^2+lr_2R_s) \sin^2 (\varphi/2) + l^2 r_2^2}}\\
  \label{eq:81}
  & \approx \frac{GM_s}{\pi R_s} \, \ln (lr_2/8R_s)\;,
\end{align}
where the integral in the first line sums up potentials of
the ring pieces at different angles $\varphi$. Translating  the
asymptotics~(\ref{eq:80}) of $U_2'$ to physical units and substituting
it into Eqs.~(\ref{eq:75}), (\ref{eq:67}), we obtain another
expression for $\Phi_s$ which should coincide with
Eq.~(\ref{eq:81}). This gives,
\begin{equation}
  \label{eq:82}
  \omega_s = \frac{l^2}{2mR_s^2} + \frac{mGM_s}{2\pi R_s}\, \ln \left(
  \beta \alpha_l / l^2 \right)\;,
\end{equation}
where we traded $R_s$ for $\alpha_l$ using Eq.~(\ref{eq:8}).  

Finally, let us compute the energy $E_s$ of the three--dimensional
Bose star at large $l$. Its analog in two dimensions has the form, 
\begin{align}
  \notag
  {\cal E}_2 & = \int d^2 \bm{x}_2\left( \frac{|\boldsymbol{\nabla_2} \psi_2|^2}{2m}
  + \frac{m}{2}\,U_2 |\psi_2|^2\right) \\
  \label{eq:76}
  &= K_2 + P_2. 
\end{align}
where $K_2$ and $P_2$ are the kinetic and potential terms,
respectively. The two--dimensional Bose star ${\psi_2 =
  \psi_{2,\, s}(\bm{x}_2)}$ 
extremizes this functional in the class of configurations with a given 
mass $M_2 \equiv \int d^2\bm{x}_2 \, m|\psi_2|^2$,
cf.\ Sec.~\ref{sec:rotating-bose-stars}. In particular, rescaling
\begin{equation}
  \label{eq:6}
  \psi_2'' = \gamma \psi_{2,s} (\gamma \bm{x}_2) \,, \;\;  U_2'' =
  U_{2,\, s} (\gamma \bm{x}_2) - 2G M_{2,s} \ln\gamma,
\end{equation}
does not change the mass and large $r_2$ asymptotics of the
potential $U_2$. Thus, the energy (\ref{eq:76}) of the rescaled
configuration 
\begin{equation}
  \label{eq:77}
  {\cal E}_{2}'' = \gamma^2 K_{2,\, s} + P_{2,\, s} - GM_{2,\, s}^2 \ln
  \gamma\;,
\end{equation}
is extremal at $\gamma=1$; hereafter we mark the quantities
evaluated for $\psi_2 = \psi_{2,\, s}$ with the subindex $s$. Taking the
$\gamma$ derivative, we find ${2K_{2,\, s} = G M_{2,\, s}^2}$ and
therefore the two--dimensional Virial theorem
\begin{equation}
  \label{eq:10}
  {\cal E}_{2,\, s} =K_{2,\, s}/2 = GM_{2,\, s}^2/4\;, 
\end{equation}
where the first equality is obtained by integrating Eq.~(\ref{eq:76})
by parts and using Eq.~(\ref{eq:68}).

At large $l$ and $\lambda=0$ the energy~(\ref{eq:56}) of the
three--dimensional Bose star takes the form, 
\begin{equation}
  \label{eq:83}
  E_s  = \frac{2 \pi R_s}{l^4} \;{\cal E}_{2,\, s}
  + \frac{\pi R_s}{m l^2}\left(\omega_s + \frac{l^2}{2m R_s^2} \right)
  \, M_{2,\, s}\;,
  \end{equation}
where Eqs.~(\ref{eq:75}), \eqref{eq:76} were used. We express
${\cal E}_{2,\, s}$ and~$\omega_s$ from Eqs.~(\ref{eq:10}) and
\eqref{eq:82}, write the two--dimensional mass as ${M_{2,\, s} = l^2
  M_s/(2\pi R_s)}$, and use representation~(\ref{eq:8}) of $R_s$. This
gives the energy expression~(\ref{eq:66}) from the main text.

Once this is done, equation (\ref{eq:71}) for $\alpha_l$ is obtained
by extremizing the energy with respect to this free
parameter. Formula~(\ref{eq:82}) for $\omega_s$ then reduces to
  the expression
$$
\omega_s = -\frac{3m^3 G^2 M_s^2}{8\pi^2 l^2} \, \alpha_l (\alpha_l+1)
$$
which was used in the estimate~(\ref{eq:15}). Notably, the value of
$\omega_s/m$
coincides with the derivative $dE_s/dM_s$, as it  should.  

%%%%%%%%%%%%%%%%%%%%%%%%%%%%%%%%%%%%%%
\section{Instability modes at $l\gg 1$}
\label{sec:instability-modes-at}
Now, consider exponentially growing modes in the background of
the three--dimensional, fast--rotating Bose star. In the main
text we argued that their profiles $\rho(\bm{x}_2)$, $\eta(\bm{x}_2)$,
$\delta \Phi(\bm{x}_2)$ satisfy the
two--dimensional eigenvalue problem (\ref{eq:87}). The latter is
similar to the problem describing vibrations of the
two--dimensional star, but includes an additional parameter: 
the extra--dimensional momentum $p_{\varphi}$. 

For a start, we prove the properties of Eqs.~(\ref{eq:87})
previewed in the main text. These equations constitute a Hamiltonian
system for the field $\rho(\bm{x}_2)$ and its canonical momentum
$\eta(\bm{x}_2)$. Indeed, in terms of the phase--space coordinate 
$\xi(\bm{x}_2) = (\rho,\, \eta)^T$ the equations read,  
\begin{equation}
  \label{eq:7}
  \mu_2\, \Omega\, \xi =  \begin{pmatrix}\hat{H}_\rho & 0\\
    0 & \hat{H}_\eta \end{pmatrix} 
  \xi\;, \qquad 
  \Omega = \begin{pmatrix} 0 & -1 \\ 1 & 0\end{pmatrix}\;,
\end{equation}
where $\mu_2$ replaces the time derivative, $\Omega$ is a canonical
form, and real symmetric operators $\hat{H}_\rho$ and $\hat{H}_\eta$
represent variations of the quadratic Hamiltonian. It is
well--known~\cite{Arnold} that the solutions of stationary
Hamiltonian systems like Eq.~(\ref{eq:7}) have only purely imaginary or real
eigenvalues $\mu_2$.

Let us demonstrate this explicitly. The operators in
Eqs.~(\ref{eq:87}) have forms,
\begin{equation}
  \notag
  \hat{H}_\eta = \frac{p_\varphi^2 - \Delta_2}{2m} + mU_2\;, \;\;
  \hat{H}_\rho = \hat{H}_\eta + \psi_2\, \frac{8\pi G 
    m^2}{\Delta_2 - p_\phi^2}\, \psi_2\;,
\end{equation}
where $U_2$ is the shifted background potential in
Eq.~(\ref{eq:75}). One can explicitly check that these operators
are real symmetric, i.e.\ satisfy the relations 
\begin{equation}
  \label{eq:27}
  \int d^2 \bm{x}_2\, f_1 \hat{H}_{\rho,\, \eta} f_2 =   \int d^2
  \bm{x}_2\, f_2 \hat{H}_{\rho,\, \eta} f_1 
\end{equation}
for any well--localized functions $f_1(\bm{x}_2)$ and
$f_2(\bm{x}_2)$. Besides, $\hat{H}_\eta$ is positive--definite because
its eigenvalues measure (non--negative) difference between the
energy levels in the potential $\Phi_2(\bm{x}_2)$ and its
ground--state energy~$\omega_2$, plus a positive constant
$p_\varphi^2/2m$. As a consequence, $\hat{H}_\eta$ and $\hat{H}_\rho$
can be simultaneously diagonalized by a real operator $\hat{A}$, 
\begin{equation}
  \label{eq:31}
  \hat{A} \hat{H}_\eta \hat{A}^T = \hat{I} \;, \qquad
  \hat{A}^{-1\, T} \hat{H}_\rho \hat{A}^{-1} = \mathrm{diag}(-\mu_j^2)\;,
\end{equation}
where $\hat{I}$ is a unity operator and all $\mu_j^2$ are real. Given
Eq.~(\ref{eq:31}), we construct a symplectic operator
\begin{equation}
  \notag
\hat{S} = \begin{pmatrix}
  \hat{A}^{-1} & 0 \\
  0 & \hat{A}^T
\end{pmatrix}\;,
\qquad \hat{S}^T\, \Omega\, \hat{S} = \Omega
\end{equation}
diagonalizing the entire boundary value problem (\ref{eq:7}). Indeed, the
transformed perturbation $\xi' = \hat{S}^{-1} \xi$ satisfies
the diagonal Hamiltonian equations (\ref{eq:7}), with operators
$\hat{I}$ and  $\mathrm{diag}(-\mu_j^2)$ replacing $\hat{H}_\eta$ and
$\hat{H}_\rho$. Solving the diagonal equations, one finds out that
every pair of elements in $\xi'$ represents the ``coordinate'' and
``momentum'' of an eigenmode with real $\mu_2^2 =
\mu_j^2$. Finally transforming to the original terms, one obtains
a complete set of eigenmodes satisfying Eqs.~(\ref{eq:87}). The
  latter modes are real by construction and their eigenvalues $\mu_2$
are either real or imaginary. 

It is clear that all modes located far away from the two--dimensional
Bose star have imaginary $\mu_2$. In this region $\psi_2\approx 0$,
and therefore
\[
  -\mu_2^2 \,\eta \approx \left[(p_\varphi^2 - \Delta_2)/2m +
    mU_2\right]^2\eta \;,
\]
see Eqs.~(\ref{eq:87}). Hence, $\mu_2^2 < 0$. Conversely,  all modes
with real $\mu_2$ are localized within the domain of nonzero $\psi_2$.

To sum up, we proved that all instability modes have real profiles
satisfying the falloff conditions~(\ref{eq:89}) at infinity. They can
be numerically computed using the standard shooting method. Performing
the rescaling with the parameter $v_0'$, we get rid of  $m$ and $G$, see
Sec.~\ref{sec:bose-stars} for details. Since the two--dimensional
background depends only on the radius  $r_2' = |\bm{x}_2'|$, we assume
generic separable dependence of perturbations on the angular
coordinate $\varphi_2 \equiv \mathrm{arctan}(y_2'/x_2')$,
\[
\rho',\, \eta',\, \delta \Phi' \propto \cos(l_2 \varphi_2 + \mbox{const})\;.
\]
This turns Eqs.~(\ref{eq:87}) into a set of ordinary differential equations
\begin{align}
  \notag
  & \partial_{r_2'} (r_2' \partial_{r_2'} \rho') = 2\mu_2' r_2' \eta' +
  \rho' \left[r_2' (p_\varphi')^2 + l_2^2/r_2' + 2 U_2' r_2'\right]\\
  \notag
  & \qquad \qquad \qquad \; + 2r_2'\psi_2' \delta \Phi'\;,\\
  \notag
  &\partial_{r_2'}(r_2' \partial_{r_2'} \eta')  = -2 \mu_2' r_2' \rho'
  + \eta' [r_2'(p_\varphi')^2 + l_2^2/r_2' + 2 U_2' r_2'
    ],\\
  \notag
  & \partial_{r_2'} (r_2' \partial_{r_2'} \delta \Phi') = \delta
  \Phi' \left[r_2' (p_\varphi')^2 + l_2^2/r_2' \right] + 8\pi r_2'
  \psi_2' \rho'
\end{align}
for the radial mode profiles $\Xi(r_2') = ( \rho',\, \eta' ,\, \delta
\Phi')$ in dimensionless units with primes.

\begin{figure}
  \centerline{\includegraphics{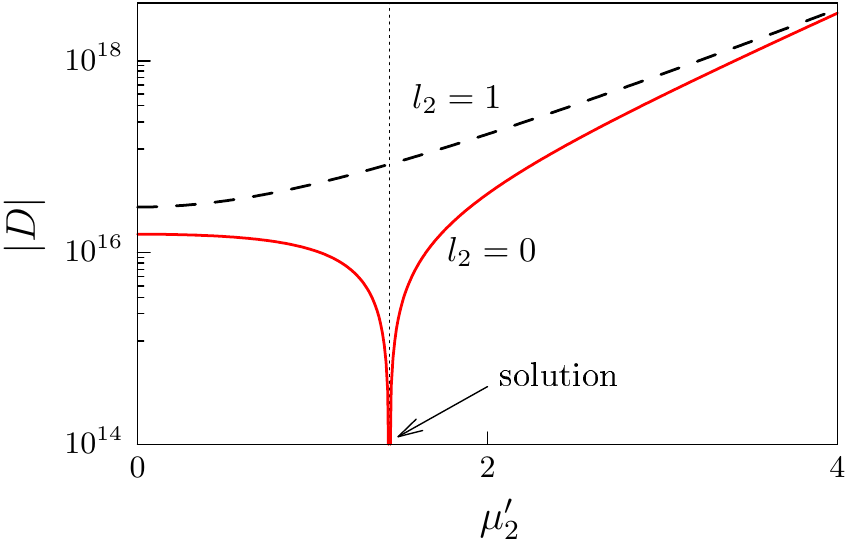}}
  \caption{Absolute value of the determinant~(\ref{eq:1}) (logarithmic
    scale) as a function of the eigenvalue $\mu_2'$ at $p_\varphi'  = 1$,
    $r_{\max}' = 7$, and $l_2 = 0,\, 1$. The solution $D=0$ (vertical
    dotted line) corresponds to a sharp beak--like dip of the $l_2 =
    0$ graph.}
  \label{fig:det}
\end{figure}

Next, we solve the equations with reguilarity conditions at the
origin,
\begin{equation}
  \label{eq:32}
  \partial_{r_2'} \Xi = 0 \qquad \mbox{at} \qquad r_2' = 0\;,
\end{equation}
and falloff conditions (\ref{eq:89}) at infinity. To this end we
construct a complete set of initial data by adding to
Eq.~(\ref{eq:32}) the condition $\Xi = (1,\, 0,\, 0)$ at $r_2' = 0$,
and then do the same for $\Xi =(0,\, 1,\, 0)$ and $(0,\, 0,\,
1)$. Starting with these three sets of data, we numerically obtain three linearly 
independent solutions $\Xi^{(1)}(r_2')$, $\Xi^{(2)}(r_2')$, and
$\Xi^{(3)}(r_2')$ of the 
differential equations. General solution is their linear
combination:
\begin{equation}
  \label{eq:33}
  \Xi(r_2') = d_1\,  \Xi^{(1)}(r_2') + d_2\, \Xi^{(2)}(r_2') + d_3\,
  \Xi^{(3)}(r_2') \;.
\end{equation}
The unknown instability mode is given by the combination satisfying,
in addition, the falloff conditions~(\ref{eq:89}) at infinity~--- or,
in numerical approximation, equalities $\Xi(r_{\max}') = 0$
at sufficiently large radius $r_2' = r_{\max}'$. Together with
Eq.~(\ref{eq:33}) this gives a system of linear algebraic equations
for $d_i$ with zero right--hand side. The solution exists only if the
coefficient matrix has zero determinant,  
\begin{equation}
  \label{eq:1}
  D = \det \left[\Xi^{(1)}\Xi^{(2)} \Xi^{(3)} \right] = 0 \qquad
  \mbox{at}\;\; r_2' = r_{\max}'\;,
\end{equation}
where the $3\times 3$ matrix under the determinant includes 
the columns $\Xi^{(i)}(r_{\max}')$. Equation (\ref{eq:1}) selects the
values of $\mu_2'$  representing the instability modes.

In Fig.~\ref{fig:det} we show the absolute value of the determinant
$D$ as a function of $\mu_2'$ at $r'_{\max} = 7 \gg 1$ and
  $p_\varphi' = 1$. We consider the cases $l_2 =0$ and $1$ (solid and
dashed lines, respectively). The sharp dip of the $l_2=0$ graph
at $\mu_2' \approx 1.44$ indicates the  point $D=0$ and thus the
instability mode. At the same time, graphs with $l_2 \geq 1$ do not
have zeros of $D$ at all, see the dashed line. This means that the
unstable mode is rotationally symmetric in two dimensions, like we
claimed in the main text.

Solving Eq.~(\ref{eq:1}) at all possible values of $p_\varphi'$ and
${l_2 = 0}$, we obtain the eigenvalue $\mu_2'(p_\varphi')$ of the
instability mode shown in Fig.~\ref{fig:pz2graph}. The maximum of this
graph represents the fastest--growing mode considered in the main
text.

%%%%%%%%%%%%%%%%%%%%%%%%%%%%%%%%%%%%%%%%%

\bibliography{rotating} 

%%%%%%%%%%%%%%%%%%%%%%%%%%%%%%%%%%%%%%%%%

\end{document}